\documentclass[fleqn,usenatbib]{mnras}

\usepackage{newtxtext,newtxmath}

\usepackage[T1]{fontenc}

\DeclareRobustCommand{\VAN}[3]{#2}
\let\VANthebibliography\thebibliography
\def\thebibliography{\DeclareRobustCommand{\VAN}[3]{##3}\VANthebibliography}

\usepackage{graphicx}
\usepackage{amsmath}

\usepackage[inline]{enumitem}
\setlist[enumerate]{labelindent=-5pt,leftmargin=*,labelsep=0.5em}
\usepackage{amsmath}
\renewcommand{\vec}[1]{{\mathbf{#1}}}
\usepackage{multirow}
\usepackage{comment}
\usepackage{placeins}

\title[Hot Jupiters in M67]{Flyby-induced high-eccentricity migration and the prevalence of hot Jupiters in M67}

\author[M. V. Kontiainen et al.]{
Mika V. Kontiainen,$^{1}$\thanks{E-mail: mvk26@cam.ac.uk (MVK)}
Cathie J. Clarke,$^{1}$
and Andrew J. Winter$^{2}$
\\
$^{1}$Institute of Astronomy, University of Cambridge, Madingley Road, Cambridge CB3 0HA, UK\\
$^{2}$Astronomy Unit, Department of Physics and Astronomy, Queen Mary University of London, Mile End Road, London E1 4NS, UK
}

\date{Accepted XXX. Received YYY; in original form ZZZ}

\pubyear{\the\year{}}

\begin{document}
\label{firstpage}
\pagerange{\pageref{firstpage}--\pageref{lastpage}}
\maketitle

\begin{abstract}
Few planetary systems form in isolation. Rather, interactions in their birth environments may sculpt their architectures or drive them into unstable configurations. Here we study the role of environmental perturbations in triggering dynamical instabilities leading to hot Jupiter formation in dense clusters, focusing on whether the elevated occurrence rate of hot Jupiters in the open cluster M67 can be explained through flyby-induced high-eccentricity migration. We develop a hybrid method for modelling the secular and tidal evolution of planetary systems under external perturbations by passing stars using a combination of analytic and numerical approaches. We evolve 10,000 realizations each of systems with either a single planet, two planets, or a planet and a stellar companion for $\sim$4 Gyr in an M67-like cluster, comparing outcome statistics against a control sample without flybys. In single- and two-planet systems, the rate of flyby-induced hot Jupiter formation is negligible. However, in systems with an initially isotropically oriented stellar companion, the cluster environment boosts hot Jupiter formation by a factor of $\sim$2, accompanied by a factor of $\sim$3 increase in the rate of planets undergoing tidal disruption. Based on observationally motivated estimates of primordial populations, our hot Jupiter yields are consistent with both field and M67 occurrence rates, provided the primordial wide binary fraction among solar-type stars is close to $\sim$50 per cent, binary-driven high-eccentricity migration is the dominant formation pathway, and a substantial fraction of tidally disrupted systems survive as hot Jupiters through partial mass loss.
\end{abstract}


\section{Introduction}

Hot Jupiters (HJs; \citealt{mayor_jupiter-mass_1995}) are a class of close-in ($P \lesssim 10$~d) giant ($m_{\rm p} \gtrsim 0.3~{\rm M_{\rm J}}$) planets with equilibrium temperatures exceeding 1,000~K. While the formation of these extreme planets at their present-day orbital radii may be possible \citep{batygin_situ_2016, boley_situ_2016}, conditions in the primordial disc generally disfavour in-situ formation, suggesting hot Jupiters likely form beyond the snowline and subsequently migrate inward either due to gas drag in the circumstellar disc \citep{lin_orbital_1996} or through high-eccentricity tidal migration (HEM; \citealt{dawson_origins_2018}).

In the HEM scenario, a cold Jupiter (CJ) on an initially wide ($a \gtrsim 1$~AU), circular ($e \sim 0$) orbit is driven to a sufficiently high eccentricity for strong tidal interactions at periapsis to shrink and circularize the orbit over time \citep{fabrycky_shrinking_2007}. As opposed to the more quiescent disc migration mechanism, HEM tends to leave hot Jupiters on orbits misaligned with respect to the spin axes of their host stars. Altogether, the paucity of young HJs \citep{karalis_separating_2025}, the discovery of highly eccentric HJ progenitors \citep{naef_hd_2001, gupta_hot-jupiter_2024}, and the observed obliquity distribution of HJ systems \citep{rice_origins_2022, spalding_tidal_2022} point to late ($\gtrsim$100~Myr) arrival via HEM as the dominant formation pathway.

While interactions with disc material can torque planets, driving them onto moderately eccentric orbits ($e \sim 0.1$; \citealt{ragusa_eccentricity_2018}), planets are unlikely to form on highly eccentric or misaligned orbits. The unstable configurations for HJ formation must therefore predominantly be triggered through post-disc interactions. In single-planet systems, proto-HJs can only be excited to high eccentricities through direct scattering by external perturbations, most notably stellar flybys \citep{zakamska_excitation_2004}. In contrast, in multi-planet systems, the eccentricity may be excited through various internal dynamical mechanisms, including planet-planet scattering \citep{rasio_dynamical_1996, chatterjee_dynamical_2008, ford_origins_2008, beauge_multiple-planet_2012} and secular chaos \citep{wu_secular_2011, hamers_secular_2017, teyssandier_formation_2019}. The presence of an outer planetary \citep{naoz_hot_2011} or stellar companion \citep{wu_planet_2003, fabrycky_shrinking_2007, anderson_formation_2016} on an inclined orbit may also drive von Zeipel-Lidov-Kozai (ZLK) oscillations \citep{von_zeipel_sur_1910, kozai_secular_1962, lidov_evolution_1962, naoz_eccentric_2016} in which the inner planet's eccentricity and inclination undergo coupled oscillations.

The prevalence of various proposed formation channels makes HJs an interesting probe of the dynamical processing undergone by planets after formation. Most stars are born in cluster environments \citep{lada_embedded_2003} in which interactions with passing stars can significantly impact the stability and dynamical outcome of planetary systems, including our solar system \citep{batygin_dynamics_2020}. As these clusters disperse, their former members populate the Galactic field. As such, interactions in the parental cluster can leave long-lasting dynamical signatures in the orbits of field planetary systems \citep{cai_signatures_2018}. Understanding this dynamical processing is crucial for constraining the outcome of the initial planet formation process \citep{ndugu_planet_2022}. Comparing the properties and frequency of HJ-hosting systems in dense clusters ($n_\star \gtrsim 100~{\rm pc^{-3}}$) and the Galactic field ($n_\star \sim 0.1~{\rm pc^{-3}}$) can therefore yield insights into the role of the stellar environment in sculpting the present-day exoplanet population.

The Galactic open cluster M67 (NGC 2682) offers a particularly well-characterized laboratory for such a comparison. It is an old ($\sim$4~Gyr; \citealt{reyes_isochrone_2024}), near-solar-metallicity ([Fe/H] $= 0.08 \pm 0.03$; \citealt{stello_k2_2016}) cluster with a present-day mass of $\sim$2,100~$\rm M_\odot$ \citep{geller_stellar_2015}, half-mass radius $\sim$3.4~pc \citep{fan_deep_1996}, and tidal radius $\sim$15.9~pc \citep{gao_machine-learning-based_2018}. M67 is one of the oldest known open clusters and is in an advanced state of dynamical evolution, having lost 60--90~per cent of its birth mass based on dynamical models \citep{hurley_complete_2005, alvarez-baena_longevity_2024}. The cluster also exhibits significant mass segregation \citep{gao_machine-learning-based_2018, carrera_extended_2019} and has a binary fraction around 25--35~per cent \citep{geller_stellar_2021, albrow_frequency_2022, wallace_photometric_2024, malhotra_stellar_2026}, substantially higher than observed in the solar neighbourhood.

In addition to being a reference cluster for the study of stellar evolution, M67 has also been targeted for planet searches. Notably, in a radial velocity (RV) survey targeting main sequence and turnoff stars in the cluster, \citet{brucalassi_search_2016} discovered three hot Jupiters. Given the limited size of the target sample, this translates to an occurrence rate significantly higher than the $\sim$1~per cent rate observed in the field in both RV \citep{mayor_harps_2011, wright_frequency_2012, wittenmyer_pan-pacific_2020} and transit surveys \citep{beleznay_exploring_2022} after correcting for binarity. In an extension to the original survey, \citet{thomas_search_2024} report revised hot Jupiter occurrence rates of $4.2_{-2.3}^{+4.1}$~per cent around all stars and $5.4_{-3.0}^{+5.1}$~per cent around single stars in M67, both implying at least a twofold increase in HJ occurrence in the cluster compared to the Galactic field.

This observation hints at an environmental correlation with hot Jupiter occurrence and raises the question whether the connection arises from 1) enhanced formation of proto-HJs due to the local chemical environment, or 2) a more efficient conversion of CJs to HJs via dynamical instabilities triggered by interactions with the crowded environment. While the giant planet occurrence rate is known to correlate strongly with host star metallicity \citep{fischer_planet-metallicity_2005, johnson_giant_2010}, the HJ progenitor fraction in M67 is unlikely to significantly exceed the field rate given the near-solar metallicity of the cluster. This points to dynamical interactions, and in particular the high binary fraction observed in M67, as the more likely driver of the elevated occurrence rate.

The direct excitation of high eccentricities by close stellar encounters in clusters was first considered by \citet{zakamska_excitation_2004}. Stellar flybys may also trigger internal instabilities in multi-planet systems by driving planets onto crossing orbits or exciting their mutual inclination, thereby activating ZLK cycles in an initially coplanar system \citep{wang_hot_2020, rodet_correlation_2021}. Recent studies have found general agreement regarding the significant role of the stellar environment in the breakdown of planetary systems \citep{cai_stability_2017, rickman_breakdown_2023} and the opening of HJ formation pathways inaccessible to isolated systems \citep{shara_dynamical_2016, hamers_hot_2017}. However, due to computational constraints, previous work has been largely limited to considering only small sample sizes or specific initial conditions, with results often extrapolated from highly disruptive flybys, neglecting the cumulative effect of many weaker perturbations which can result in a random walk in eccentricity and inclination.

Recently, \citet{wirth_hot_2025} developed an efficient hybrid method for modelling the long-term evolution of single-planet systems in cluster environments. The efficiency of their model is based on employing the analytic eccentricity excitation equations derived by \citet{heggie_effect_1996} for flybys in the secular (tidal, slow) regime and only resorting to direct $N$-body simulations when these assumptions break down. As a case study, they applied their hybrid model to the evolution of stellar systems hosting a single Jupiter-mass planet over 12~Gyr in a toy model of the globular cluster 47 Tuc based on the results of \citet{giersz_monte_2011}, finding a roughly twofold increase in the production of hot Jupiters compared to purely analytic models \citep{winter_forming_2022-1}.

In this work, we extend this hybrid framework to planetary systems containing an outer companion (planet or star), including an analytic prescription for inclination excitation following \citet{rodet_odea_2019}. Given that $N$-body integrations over Gyr timescales are computationally prohibitive, we follow a similar hybrid approach in modelling the orbital evolution of the planetary system between flybys. Specifically, we utilize the orbit-averaged secular evolution equations whenever they are valid, only resorting to a direct $N$-body approach during periods of strong interaction. This approach allows us to efficiently trace the complete dynamical history of planetary systems in cluster environments over long timescales. To demonstrate our methodology, we apply our model to investigate HJ formation in an M67-like cluster and compare our results to previous work on the topic.

The paper is structured as follows. In Section \ref{sec:methods}, we present the core aspects of the dynamical model with application to M67. We present our results in Section \ref{sec:results} and discuss their implications in the context of both observations and previous theoretical work in Section \ref{sec:discussion}. We conclude in Section \ref{sec:conclusions} with a summary of our findings.

\section{Methods}
\label{sec:methods}

\subsection{Setup}

\subsubsection{Star Cluster}
\label{sec:starcluster}

\begin{figure}
\includegraphics[width=\columnwidth]{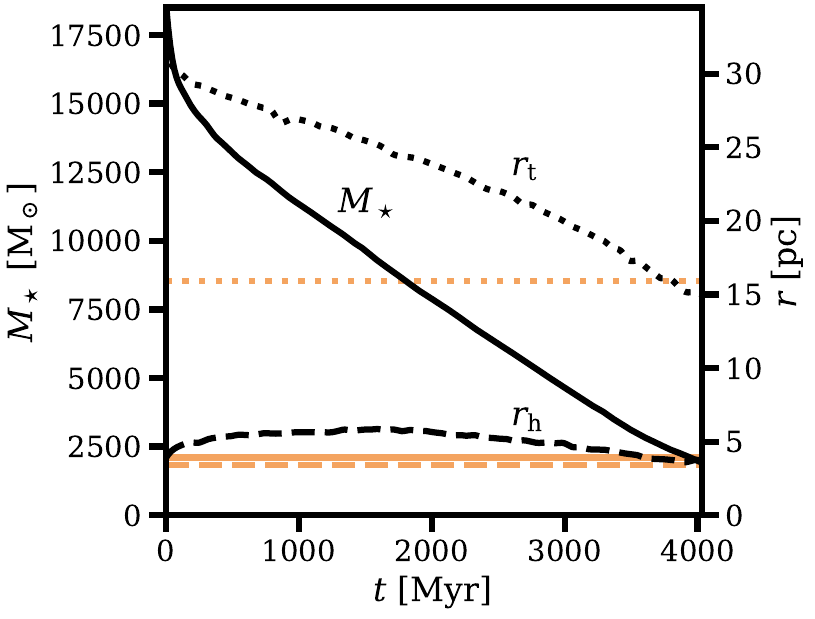}
\caption{Evolution of the total cluster mass $M_\star$ (solid), half-mass radius $r_{\rm h}$ (dashed), and tidal radius $r_{\rm t}$ (dotted) based on the \citet{hurley_complete_2005} model. The final values at $t \sim 4$~Gyr are in good agreement with the corresponding observed properties of M67 shown as orange horizontal lines.
\label{fig:cluster_evolution}}
\end{figure}

\begin{figure}
\includegraphics[width=\columnwidth]{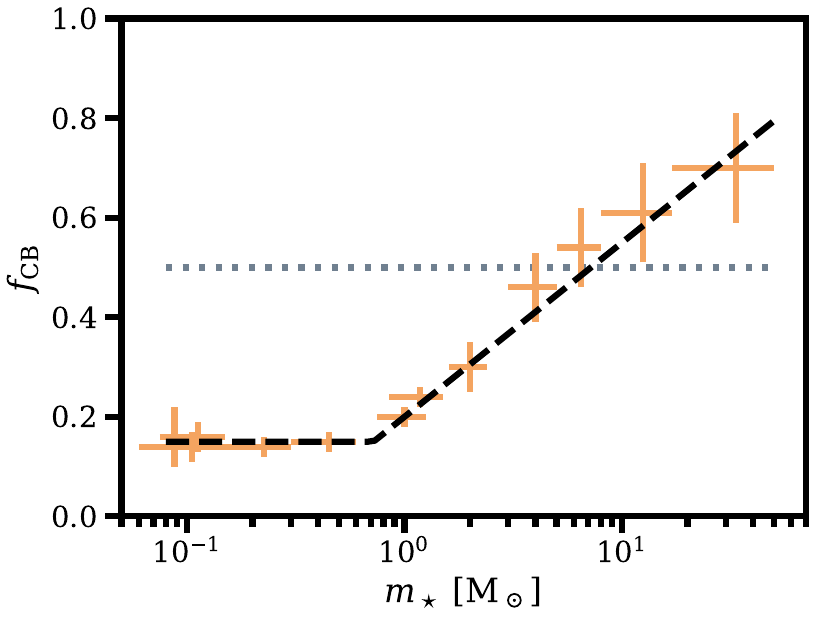}
\caption{The close ($a_{\rm bin} < 10$~AU) binary fraction as a function of primary mass. The orange data points are taken from Table 1 of \citet{offner_origin_2023} and fit with the black dashed line (Equation \ref{eq:cbf}). For reference, the constant binary fraction of 0.5 used for an additional experiment presented in Appendix \ref{sec:binaryfraction} is also shown as a grey dotted line.
\label{fig:binary_fraction}}
\end{figure}

The computational expense of full $N$-body models of dense clusters (e.g. \citealt{heggie_towards_2014}) precludes their use in population synthesis models. To incorporate the effects of background cluster evolution in a computationally efficient way, we follow \citet{wirth_hot_2025} in treating the cluster as an evolving Plummer sphere \citep{plummer_problem_1911} with the potential
\begin{equation}
    \Phi(r, t) = - \frac{G M_\star(t)}{\sqrt{r^2 + a(t)^2}}
\end{equation}
where the Plummer scale parameter $a$ is related to the instantaneous half-mass radius $r_{\rm h}$ by $a \approx 0.766~r_{\rm h}$.

For a Plummer sphere, the local stellar mass density and 1D velocity dispersion at a given time $t$ and radius $r$ from the cluster core are given by
\begin{equation}
    \rho_\star(r, t) = \frac{3 M_\star(t)}{4 \pi a(t)^3} \left( 1 + \frac{r^2}{a(t)^2} \right)^{-5/2}
\end{equation}
and
\begin{equation}
    \sigma_{v}^2 (r, t) = \frac{G M_\star(t)}{6 \sqrt{r^2 + a(t)^2}}
\end{equation}
where the global properties (total stellar mass $M_\star$, half-mass radius $r_{\rm h}$, and tidal radius $r_{\rm t}$) of the cluster evolve with time following prescribed evolutionary tracks. For our case study of M67, we base the cluster evolution on the $N$-body model of \citet{hurley_complete_2005}, global snapshots of which are available at $\sim$60~Myr intervals.\footnote{Snapshots of the \citet{hurley_complete_2005} $N$-body model of M67 are available at \href{https://astronomy.swin.edu.au/~jhurley/nbody/archive.html}{https://astronomy.swin.edu.au/~jhurley/nbody/archive.html}.} Between these, the values of the global cluster parameters are interpolated using a cubic spline. The evolution of the main cluster parameters is shown in Figure \ref{fig:cluster_evolution}. The final properties of the simulated cluster provide a good match to the observed mass (${\sim} 2,100~{\rm M_\odot}$; \citealt{geller_stellar_2015}), half-mass radius (${\sim} 3.4~{\rm pc}$; \citealt{fan_deep_1996}), and tidal radius (${\sim} 15.9~{\rm pc}$; \citealt{gao_machine-learning-based_2018}) of M67.

In this work, we consider only host stars of mass $m_{\rm s} = 1~{\rm M_\odot}$, the initial positions $\vec{x}_{\rm s}$ and velocities $\vec{v}_{\rm s}$ of which are sampled from a Plummer sphere in virial equilibrium following the algorithm presented in \citet{aarseth_comparison_1974}. The mass distribution of other cluster member stars is taken to be a broken \citet{kroupa_variation_2001} initial mass function (IMF) of the form
\begin{equation}
\label{eq:imf}
    \xi(m_\star) \propto
    \begin{cases}
        m_\star^{-1.3}, & 0.08~{\rm M_\odot} \leq m_\star < 0.5~{\rm M_\odot} \\
        m_\star^{-2.3}, & 0.5~{\rm M_\odot} \leq m_\star \leq 50~{\rm M_\odot}
    \end{cases}
\end{equation}
which \citet{hunt_improving_2024} found to be consistent with the Galactic open cluster population. This corresponds to a mean stellar mass $\langle m_\star \rangle \approx 0.55~{\rm M_\odot}$. We also adopt a mass-dependent close ($a_{\rm bin} < 10$~AU) binary fraction based on the values compiled in Table 1 of \citet{offner_origin_2023}. We fit these values with the following function:
\begin{equation}
\label{eq:cbf}
    f_{\rm CB}(m_\star) = \max{\left\{f_{\rm CB,\:floor},\: f_{\rm CB,\:1~M_\odot} + \alpha \log_{10}\left( \frac{m_\star}{1~\rm M_\odot} \right) \right\}}
\end{equation}
where the low-mass binary fraction $f_{\rm CB,\:floor} = 0.15$, the solar-mass binary fraction $f_{\rm CB,\:1~M_\odot} = 0.2$, and the slope $\alpha = 0.35$. This function provides a good fit to the \citet{offner_origin_2023} values (Figure \ref{fig:binary_fraction}) and is in fair agreement with observational surveys which find 25--35~per cent of main sequence cluster members in M67 to be in binaries \citep{geller_stellar_2021, albrow_frequency_2022, wallace_photometric_2024, malhotra_stellar_2026}.\footnote{To investigate the role of the binary fraction in more detail, we ran an additional experiment using a fixed binary fraction of 0.5, the results of which are discussed in Appendix \ref{sec:binaryfraction}.} We further adopt a uniform binary mass ratio $q$ distribution between 0.1--1 such that $\langle q \rangle = 0.55$. With these choices, the mean mass of stellar systems in the cluster becomes $\langle m_{\rm sys} \rangle \approx 0.66~{\rm M_\odot}$ after correcting for binarity.

\subsubsection{Planetary System}
\label{sec:planetarysystem}

\begin{figure}
\includegraphics[width=\columnwidth]{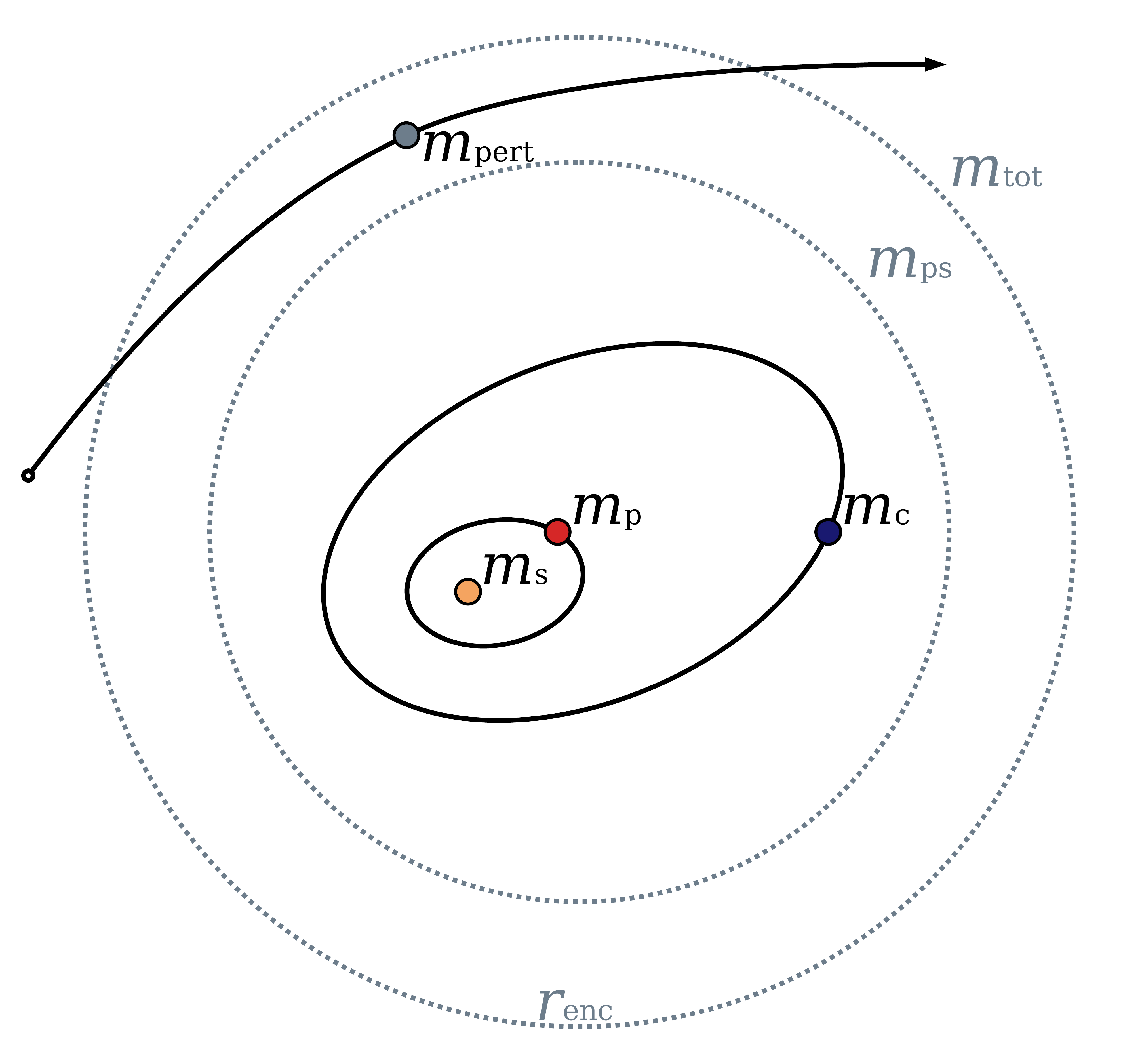}
\caption{General configuration of the planetary system consisting of the host star (orange), a planet (red), an outer companion (blue), and a temporary stellar perturber (grey) passing through the encounter sphere of radius $r_{\rm enc}$. Encounters within this radius are modelled explicitly, whereas more distant flybys outside the sphere are treated as unresolved background perturbations using the \citet{chandrasekhar_dynamical_1943} formalism (Section \ref{sec:stellardynamics}).
\label{fig:psystem}}
\end{figure}

We consider planetary systems consisting of a 1~$\rm M_\odot$ host star orbited by a Jovian planet with an outer planetary or stellar companion (Figure \ref{fig:psystem}). Specifically, the three initial configurations we consider are as follows:
\begin{enumerate}
    \item \textbf{Single Planet (1P)}: A Jovian planet with mass $m_{\rm p} = 1~{\rm M_{J}}$ and semi-major axis $a_{\rm p}$ sampled log uniformly between 1--10~AU.
    \item \textbf{Planet + Planetary Companion (2P)}: Similar to Case (i) but with the addition of an outer planetary companion with mass $m_{\rm c}$ sampled uniformly between 1--13~$\rm M_{J}$ and semi-major axis $a_{\rm c}$ log uniformly between 1--100~AU subject to $a_{\rm c} > a_{\rm p}$.
    \item \textbf{Planet + Stellar Companion (1P1S)}: Similar to Case (i) but with the addition of an outer stellar companion with mass $m_{\rm c}$ sampled uniformly between 0.1--1~$\rm M_\odot$ and semi-major axis $a_{\rm c}$ log uniformly between 100--1,000~AU.
\end{enumerate}

Our choice of sampling the semi-major axis of the inner planet is motivated by the findings of \citet{nielsen_gemini_2019} and \citet{fernandes_hints_2019}. We choose the maximum orbital radius of planetary companions based on more distant companions being readily disrupted by individual encounters \citep{carter_survivability_2023}. However, given the prevalence of disc truncation in cluster environments \citep{winter_protoplanetary_2018, parker_birth_2020}, this limit may be regarded as an optimistic estimate. In particular, observations of the young ($\sim$1~Myr) Orion Nebula Cluster suggest protoplanetary discs in dense environments do not generally exceed 30~AU in radial extent \citep{eisner_protoplanetary_2018, tobin_vlaalma_2020}. The orbital radius distribution of stellar companions is similarly motivated by environmental considerations. While even wider binaries may play a significant role in the formation of hot Jupiters in the Galactic field \citep{grishin_hot_2025}, such systems are unlikely to survive for long in dense environments \citep{cournoyer-cloutier_massive_2024}. We also limit our binary mass ratio distribution to $q \leq 1$, corresponding to $m_{\rm c} < 1~{\rm M_\odot}$, noting that 91~per cent of all planets in multiple star systems are found orbiting the more massive primary component \citep{michel_gaia_2024}. Furthermore, as only companions with $m_{\rm c} \lesssim 1.4~{\rm M_\odot}$ are expected to remain on the main sequence for the entire duration of the simulation ($\sim$4~Gyr), a self-consistent model of $q > 1$ systems would require a detailed prescription of stellar evolution.\footnote{We note that binary-induced ZLK migration may be more efficient when the planet orbits the less massive secondary component \citep{liu_double_2026}. For such systems, the white dwarf natal kick associated with the evolution of a more massive primary off the main sequence may also open up additional HJ formation pathways \citep{stephan_two_2024}.}

In order to probe orbital parameters representative of the primordial planet population, we sample the initial planetary eccentricities from a Rayleigh distribution truncated at 0.6 and with a scale parameter 0.05. The initial orbital inclinations for planets are drawn from a normal distribution centred at 0 and with a standard deviation of 2$^\circ$. These distributions are consistent with the observed statistical properties of multi-planet systems as reported by \citet{xie_exoplanet_2016}. On the other hand, stellar companions likely form outside of the disc and therefore exhibit different orbital properties. While even relatively wide ($\sim$700~AU) binary companions appear moderately aligned with the orbits of small planets \citep{dupuy_orbital_2022, christian_possible_2022}, a similar pattern has yet to emerge for giant planets \citep{christian_wide_2025}. As such, we sample the initial inclinations of stellar companions isotropically between 0--180$^\circ$ (uniformly in $\cos{i}$) and eccentricities uniformly between 0--1 \citep{raghavan_survey_2010, moe_mind_2017}, in line with previous population synthesis studies \citep{weldon_saving_2026}. The argument of periapsis $\omega$ and the longitude of ascending node $\Omega$ are both sampled uniformly between 0--360$^\circ$. In both cases, the orbital parameters are sampled subject to the system being stable and hierarchical (secularly interacting) based on the criteria presented in Section \ref{sec:planetarydynamics}.

\subsection{Evolution between Flybys}

\subsubsection{Stellar Dynamics}
\label{sec:stellardynamics}

\begin{figure*}
\includegraphics[width=\textwidth]{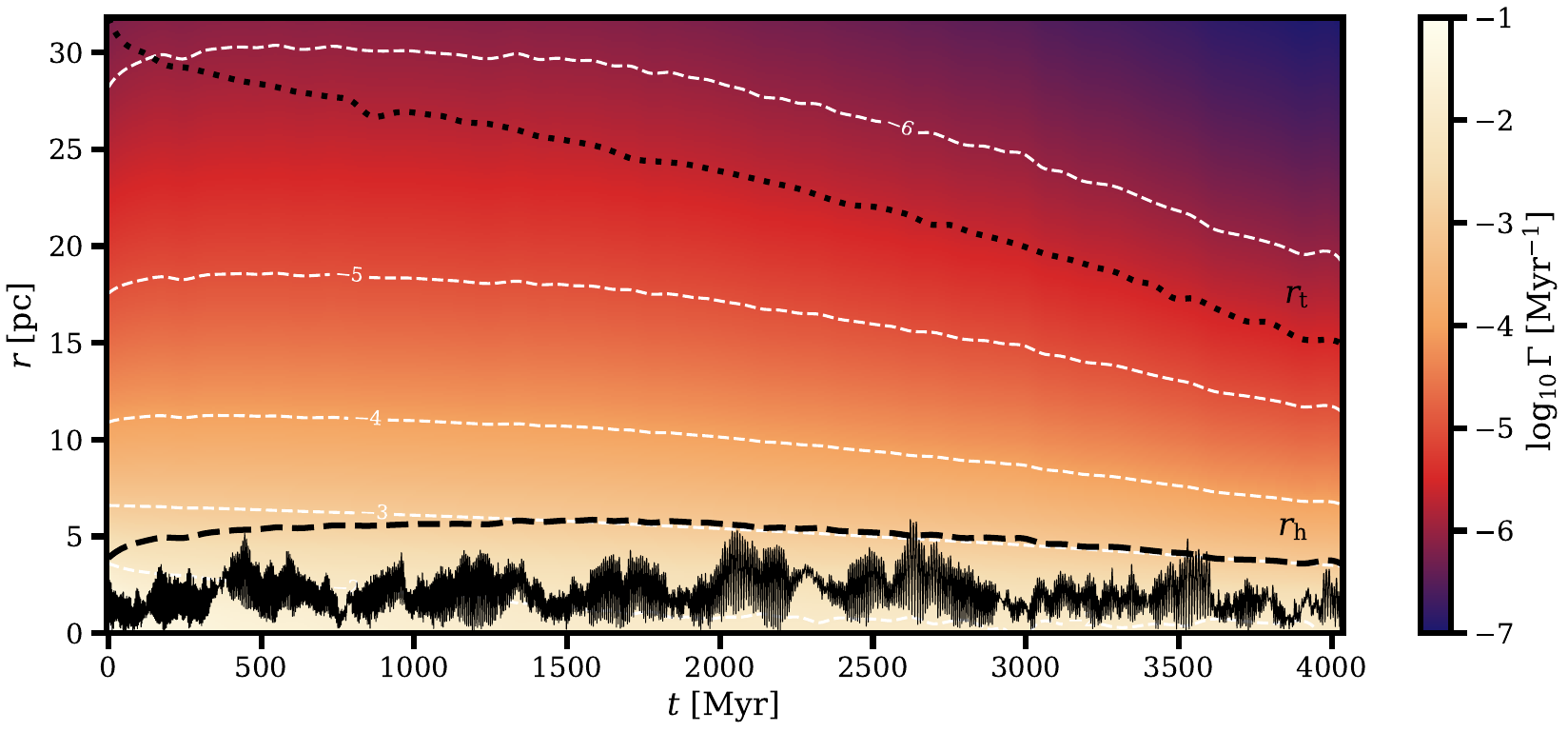}
\caption{The total instantaneous encounter rate $\Gamma$ with fixed $r_{\rm enc} = 1,000$~AU as a function of time $t$ and cluster-centric radius $r$ in our toy model of M67 using the $\sim$60~Myr snapshots from \citet{hurley_complete_2005} for the prescription of cluster evolution. The evolutionary tracks of the tidal radius $r_{\rm t}$ and half-mass radius $r_{\rm h}$ are shown in black. For reference, the stellar number density and velocity dispersion at the centre of the cluster at the beginning (end) of the simulation are 260 (35)~$\rm pc^{-3}$ and 2.1 (0.7)~$\rm km~s^{-1}$, respectively. A single trajectory of a host star that remains predominantly within the half-mass radius is shown for illustration.
\label{fig:encounter_rate}}
\end{figure*}

Unlike \citet{wirth_hot_2025} who fixed planetary systems at constant Lagrangian radii in the cluster, we treat the planetary system (host star) as a test particle and simulate its trajectory in an evolving Plummer potential using the Dormand-Prince 85(3) integrator implemented in \texttt{gala} using a time step $\delta t = 0.1~{\rm Myr}$. The simultaneous evolution of the mass and scale radius of the Plummer potential is handled using the \texttt{TimeInterpolatedPotential} class. We also incorporate the effects of dynamical friction and diffusion from weak, unresolved encounters on the stellar trajectory by applying small velocity kicks based on the \citet{chandrasekhar_dynamical_1943} formalism (Equation 7.92 of \citealt{binney_galactic_2008}) at each step. The corresponding Coulomb logarithm is evaluated dynamically as $\ln{\Lambda} = \ln{(r_{\rm t}/r_{\rm enc})}$ where $r_{\rm t}$ is the tidal radius of the cluster and $r_{\rm enc}$ is the radius of the encounter sphere, corresponding to the maximum periapsis distance below which encounters are modelled explicitly. Together with the evolutionary prescription for the background potential (Section \ref{sec:starcluster}), the velocity diffusion leads to the mass segregation and gradual dispersal of the cluster. To model the escape of stars from the cluster, we follow \citet{giersz_monte_2008} in imposing a strict tidal energy cut-off based on the cluster potential evaluated at the tidal radius $\Phi(r_{\rm t})$. To avoid the premature removal of stars that are temporarily energetically unbound but still dynamically coupled to the cluster, a host star is only considered unbound if it crosses the tidal boundary $r_{\rm t}$ and its specific orbital energy exceeds $\Phi(r_{\rm t})$. Any subsequent dynamical evolution is then modelled without flybys.

We estimate the rate of resolved close encounters using the commonly used mean free path ($n_\star \Sigma v_\infty$) approximation \citep{reinoso_mean_2022}. While the toy model of \citet{wirth_hot_2025} included only the geometric cross-section which is dominant in globular clusters with high relative velocities between stars, gravitational focusing can dramatically increase the rate of close encounters in open clusters with low velocity dispersions. We account for this effect by estimating the rate of encounters with periapsis distance within the encounter sphere ($r_{\rm close} \leq r_{\rm enc}$) using the effective cross-section
\begin{equation}
    \Sigma{(v_\infty)} = \pi r_{\rm enc}^2 \left(1 + \frac{2Gm_{\rm tot}}{r_{\rm enc} v_\infty^2} \right)
\end{equation}
where $m_{\rm tot} = m_{\rm ps} + m_{\rm pert}$ is the total mass of the planetary system and perturber, and the background velocity distribution is taken to be a Maxwellian
\begin{equation}
    f{(v_\infty)} = \sqrt{\frac{2}{\pi}} \frac{v_\infty^2}{\sigma_{\rm rel}^3} \exp{\left( -\frac{v_\infty^2}{2 \sigma_{\rm rel}^2} \right)}
\end{equation}
where $\sigma_{\rm rel} = \sqrt{2} \sigma_v$ is the local relative velocity dispersion.

The instantaneous encounter rate is then given by
\begin{equation}
\label{eq:gamma_integral}
    \Gamma = n_\star \int_0^\infty \Sigma{(v_\infty)} v_\infty f{(v_\infty)} dv_\infty
\end{equation}
where $n_\star$ is the local stellar number density. The local mass density is converted to a stellar number density via $n_\star = \rho_\star /\langle m_{\rm sys} \rangle$. Performing the integration over $v_\infty \in [0, \infty]$ in Equation \ref{eq:gamma_integral} yields the geometric and focused encounter rates:
\begin{equation}
\label{eq:gamma_geo}
    \Gamma_{\rm geo} = 2 \sqrt{2 \pi} r_{\rm enc}^2 n_\star \sigma_{\rm rel}
\end{equation}
and
\begin{equation}
\label{eq:gamma_focused}
    \Gamma_{\rm focus} = \frac{2 \sqrt{2 \pi} r_{\rm enc} G m_{\rm tot} n_\star}{\sigma_{\rm rel}}
\end{equation}
where we set the perturber mass $m_{\rm pert}$ to equal the mean mass of stellar systems in the cluster $\langle m_{\rm sys} \rangle$ in evaluating the total mass of the interacting bodies $m_{\rm tot}$.

The total encounter rate $\Gamma = \Gamma_{\rm geo} + \Gamma_{\rm focus}$ is shown as a function of time and cluster-centric radius with fixed $r_{\rm enc} = 1,000$~AU in Figure \ref{fig:encounter_rate}. While this is a typical value for the definition of a stellar flyby across previous literature \citep{pfalzner_early_2013, winter_protoplanetary_2018, cuello_close_2023}, significant efficiency and accuracy gains can be made by adjusting $r_{\rm enc}$ based on the size of the system. Since we are interested in the effect of close encounters on both the stellar trajectory and planetary orbits, we define $r_{\rm enc}$ dynamically as
\begin{equation}
\label{eq:renc}
    r_{\rm enc} = \max\left\{ r_{\rm str},\: 10 \times r_{\rm apo}^{\rm max} \right\}
\end{equation}
where $r_{\rm apo}^{\rm max} = \max\{a_{\rm p}(1+e_{\rm p}),\: a_{\rm c}(1+e_{\rm c})\}$ is the maximum apoapsis distance in the system and $r_{\rm str}$ is the strong encounter radius of the planetary system treated as a point mass:
\begin{equation}
\label{eq:rstr}
    r_{\rm str} = \frac{2 G m_{\rm tot}}{\sigma_{\rm rel}^2}
\end{equation}
corresponding to a 90$^\circ$ deflection.\footnote{In practice, $r_{\rm enc}$ is additionally capped at half the local mean interstellar spacing $\langle d_\star\rangle \sim n_\star^{-1/3}$ to preserve the isolated encounter assumption underlying the Poisson sampling framework. Were $r_{\rm enc}$ to exceed this scale, the encounter sphere would on average contain multiple perturbing stars simultaneously, violating the assumption that flybys are sequential and independent.} While the effect of cumulative small-angle deflections due to weak encounters beyond this radius is captured by the \citet{chandrasekhar_dynamical_1943} formalism, that of strong encounters with closest approach distance $r_{\rm close} < r_{\rm str}$ must be modelled explicitly. Requiring $r_{\rm enc} \geq r_{\rm str}$ therefore guarantees the effect of velocity kicks from both strong and weak encounters on the stellar trajectory are handled self-consistently. On the other hand, our choice of the factor of 10 in Equation \ref{eq:renc} reflects the need to include all relevant encounters while keeping the computational cost feasible. This choice is analogous to, though more conservative than, the threshold adopted by \citet{winter-granic_binary_2024} who found flybys with impact parameter $b \gtrsim 5 \times a$ to have negligible influence on the internal dynamics of binaries in the impulse limit. Similarly, in the context of multi-planet systems, \citet{rodet_correlation_2021} and \citet{wang_hot_2022} demonstrated that individual flybys with $r_{\rm close}$ further than $\sim$2 times the outermost orbital radius do not significantly perturb an initially coplanar, circular configuration. Consistent with these findings, we found that further extending the radius of the encounter sphere $r_{\rm enc}$ had no significant effect on the outcome statistics of our model.

Given the dependence of the encounter rate $\Gamma$ on both the cluster-centric radius $r$ and time $t$, the standard tau-leaping approach, in which the time until the next flyby is estimated based on the local waiting time $\tau \sim 1/n_\star \Sigma v_\infty$ (e.g. \citealt{winter-granic_binary_2024}), breaks down. Instead, we develop a framework for modelling the evolution of the planetary system under stochastic flybys as an inhomogeneous Poisson process using the Euler-Maruyama method \citep{higham_algorithmic_2001}. Evolving the system concurrently in $\delta t = 0.1$~Myr steps, we calculate the expected number of encounters prior to each timestep as $\lambda = \Gamma(r, t) \delta t$. We then sample the actual number of realized encounters from a Poisson distribution with an expectation value $\lambda$. In this way, the full stellar trajectory is used to inform the total number of encounters, the probability of which is naturally weighted by the properties of the local stellar environment. We demonstrate the self-consistency of this approach in Appendix \ref{sec:selfconsistency}.

\subsubsection{Planetary Dynamics}
\label{sec:planetarydynamics}

In terms of duration, close encounters constitute only a small fraction of the overall cluster lifetime, so it is crucial that the evolution of the planetary system in the intervals between flybys is appropriately captured. Given the computational expense of running long-term $N$-body simulations, we identify the planetary system as belonging to one of two possible dynamical regimes following a flyby -- secularly interacting or strongly interacting -- and adjust our approach to modelling the internal dynamics appropriately.

If the orbital configuration of system is sufficiently hierarchical, such that the two bodies are not exchanging energy through strong interactions, the individual orbits may be averaged over, resulting in the secular approximation. For this to be valid, we require the following conditions to be met:
\begin{enumerate}
    \item The system must be \textbf{Hill stable}. In the case of an outer planetary companion, the \citet{petrovich_stability_2015} stability criterion must be satisfied:
    \begin{equation}
    \label{eq:petrovich}
        \frac{a_{\rm c}(1 - e_{\rm c})}{a_{\rm p}(1 + e_{\rm p})} > 2.4 [\max{\{ \mu_{\rm p},\: \mu_{\rm c} \}}]^{1/3} \left( \frac{a_{\rm c}}{a_{\rm p}} \right)^{1/2} + 1.15.
    \end{equation}
    Alternatively, if the outer companion is a star, the \citet{mardling_tidal_2001} stability criterion must be satisfied:
    \begin{equation}
    \label{eq:mardling}
        \frac{a_{\rm c}}{a_{\rm p}} > 2.8 \left( 1 + \frac{m_{\rm c}}{m_{\rm s} + m_{\rm p}} \right)^{2/5} \frac{(1 + e_{\rm c})^{2/5}}{(1 - e_{\rm c})^{6/5}} \left( 1 - \frac{0.3i_{\rm mut}}{180^\circ} \right).
    \end{equation}
    \item The system must be \textbf{AMD stable}. The angular momentum deficit (AMD) of the system must be small enough to preclude orbit crossing, such that the AMD stability criterion of \citet{laskar_amd-stability_2017} is satisfied.
    \item The system must not be in the \textbf{semi-secular} regime. Specifically, the quadrupole timescale of angular momentum exchange between the orbits ($\tau_{\rm ZLK}$; \citealt{antognini_timescales_2015}) must exceed the outer orbital period:
    \begin{equation}
    \label{eq:quadrupole}
        \tau_{\rm ZLK} / P_{\rm c} = \frac{8}{15 \pi} \frac{m_{\rm ps} }{m_{\rm c}} \frac{P_{\rm c}}{P_{\rm p}} (1 - e_{\rm c}^2)^{3/2} > 1
    \end{equation}
    and the system must be sufficiently hierarchical, i.e. the hierarchy (octupole) parameter $\epsilon$ must be small:
    \begin{equation}
        \epsilon = \frac{a_{\rm p}}{a_{\rm c}} \frac{e_{\rm c}}{1 - e_{\rm c}^2} < 0.1.
    \end{equation}
    \item The system must not be in the proximity of a low-order \textbf{mean-motion resonance}. Near such resonances, the resonant angle may librate rather than circulate freely, violating the assumption of orbital averaging that underlies the secular equations. We identify near-resonant configurations using the chaotic zone overlap criterion of \citet{wisdom_resonance_1980}, which gives the half-width of the chaotic zone surrounding a planet of mass ratio $\mu = m_{\rm c}/m_{\rm s}$ as $(\delta a/a)_{\rm chaos} = 1.3\mu^{2/7}$. Converting to a width in period ratio space via $P \propto a^{3/2}$, we require that the period ratio $P_{\rm c}/P_{\rm p}$ does not fall within $1.5 \times 1.3\mu^{2/7}$ of any first-order commensurability $(j+1){:}j$ for $j = 1, \ldots, 5$:
    \begin{equation}
        \left| \frac{P_{\rm c}}{P_{\rm p}} - \frac{j+1}{j} \right| > 1.95\mu^{2/7} \frac{j+1}{j}.
    \end{equation}
\end{enumerate}

If these conditions are satisfied, we model the dynamical evolution of the system by numerically integrating the octupole-level secular evolution equations (Equations A1--A9 in \citealt{liu_suppression_2015}) using the \texttt{solve\_ivp} function in \texttt{SciPy} \citep{virtanen_scipy_2020}. If the mutual inclination between the planet and its companion exceeds the critical Kozai angle $\approx 39.2^\circ$, the system will undergo periodic von Zeipel-Lidov-Kozai (ZLK) oscillations which can drive the eccentricity of the inner body to extremely high values \citep{naoz_eccentric_2016}. These oscillations as well as more modest long-term variations in eccentricity and inclination are captured by our secular evolution prescription.

In the complementary regime, the two bodies may undergo strong scattering events which alter their orbital properties on orbital timescales. In such cases the system is evolved using the 15th-order adaptive integrator \texttt{IAS15} \citep{rein_ias15_2015} based on the Gauss-Radau method \citep{everhart_efficient_1985} and implemented in \texttt{REBOUND} \citep{rein_rebound_2012}. The adaptive nature of the integrator ensures close encounters between the two bodies are handled accurately without expending excessive computational resources during periods of more quiescent evolution. If mutual interactions render the system dynamically stable and only secularly interacting at any stage of the simulation, we transition back to the secular approximation.

In either regime, tides are activated when the periapsis of the planet $r_{\rm peri} < 0.5~{\rm AU}$, as larger values lead to minimal tidal evolution over the cluster lifetime. For the purposes of modelling the tidal interaction between the planet and the host star, we take their physical radii to be $R_{\rm p} = 1~{\rm R_J} \approx 4.7\times10^{-4}~{\rm AU}$ and $R_{\rm s} = 1~{\rm R_\odot} \approx 4.7\times10^{-3}~{\rm AU}$, respectively. We include both equilibrium \citep{darwin_bodily_1879, goldreich_q_1966, alexander_weak_1973, hut_tidal_1981} and dynamical tides \citep{press_formation_1977, moe_dynamical_2018} in our tidal prescription following the hybrid approach of \citet{rozner_inflated_2022}. We neglect tidal dissipation within the host star, as this is thought to dominate only once the planet's eccentricity and obliquity have been damped by planetary tides \citep{millholland_empirical_2025}.

In the equilibrium tide model \citep{darwin_bodily_1879, goldreich_q_1966, alexander_weak_1973, hut_tidal_1981}, the orbit-averaged equations for the evolution of the semi-major axis and eccentricity of a pseudo-synchronized ($\Omega \approx n$) planet are
\begin{equation}
    \frac{da}{dt} \biggr|_{\rm eq} = -21 k_{\rm p} n^2 \tau_{\rm p} \frac{m_{\rm s}}{m_{\rm p}} \left( \frac{R_{\rm p}}{a} \right)^5 a e^2 \frac{f(e)}{(1 - e^2)^{15/2}},
\end{equation}
\begin{equation}
    \frac{de}{dt} \biggr|_{\rm eq} = -\frac{21}{2} k_{\rm p} n^2 \tau_{\rm p} \frac{m_{\rm s}}{m_{\rm p}} \left( \frac{R_{\rm p}}{a} \right)^5 e \frac{f(e)}{(1 - e^2)^{13/2}},
\end{equation}
assuming angular momentum is conserved during migration. Here $k_{\rm p} = 0.25$ is the apsidal motion constant, $n$ is the mean motion of the planet, and $\tau_{\rm p} = 0.66$~s is the planetary tidal lag \citep{hamers_hot_2017}. The function $f(e)$ is defined as
\begin{equation}
    f(e) = \frac{1 + \frac{45}{14}e^2 + 8e^4 + \frac{685}{224}e^6 + \frac{255}{448}e^8 + \frac{25}{1792}e^{10}}{1 + 3e^2 + \frac{3}{8}e^4}.
\end{equation}

During periods of very high eccentricity, the excitation of internal energy modes of the planet may result in an enhanced tidal response. At the quadrupole order, the energy dissipated during a single periapsis passage is given by
\begin{equation}
    \Delta E = f_{\rm dyn} \frac{m_{\rm s} + m_{\rm p}}{m_{\rm p}} \frac{G m_{\rm s}^2}{R_{\rm p}} \left( \frac{a(1 - e)}{R_{\rm p}} \right)^{-9}
\end{equation}
where the efficiency of energy deposition is taken to be $f_{\rm dyn} = 0.1$ for gas giants \citep{press_formation_1977, moe_dynamical_2018}. Assuming a constant periapsis distance during migration, the semi-major axis and eccentricity evolution are described by
\begin{equation}
    \frac{da}{dt} \biggr|_{\rm dyn} = \frac{a}{P} \frac{\Delta E}{E},
\end{equation}
\begin{equation}
    \frac{de}{dt} \biggr|_{\rm dyn} = \frac{1 - e}{a} \frac{da}{dt},
\end{equation}
where $P$ is the orbital period of the planet. Following \citet{rozner_inflated_2022}, we switch between equilibrium and dynamical tides based on the ratio of the migration rates
\begin{equation}
    \beta{(a, e)} \coloneq \frac{da/dt |_{\rm dyn}}{da/dt |_{\rm eq}} = \frac{2 f_{\rm dyn} R_{\rm p}^3 (1 - e^2)^{15/2}}{21 G m_{\rm p} k_{\rm p} \tau_{\rm p} (1 - e)^9 P e^2 f(e)}
\end{equation}
such that dynamical tides are activated when $\beta > 1$ and $e > 0.2$. The lower eccentricity cutoff is imposed to avoid the divergence of dynamical tides at small eccentricities.

Finally, we also include apsidal precession due to tides and general relativity (but not due to oblateness/rotation) with the respective precession rates given by:
\begin{equation}
\begin{split}
    \dot \omega_{\rm tide} = &\frac{15 G^{1/2} (m_{\rm s} + m_{\rm p}) ^{1/2}}{8 a^{13/2}} \frac{8 + 12e^2 + e^4}{(1 - e^2)^5} \\ &\times \left[ \frac{m_{\rm p}}{m_{\rm s}} k_{\rm s} R_{\rm s}^5 + \frac{m_{\rm s}}{m_{\rm p}} k_{\rm p} R_{\rm p}^5 \right]
\end{split}
\end{equation}
and
\begin{equation}
    \dot \omega_{\rm GR} = \frac{3 G^{3/2} (m_{\rm s} + m_{\rm p}) ^{3/2}}{a^{5/2} c^2 (1 - e^2)}
\end{equation}
where $k_{\rm s} = 0.014$ and $k_{\rm p} = 0.25$ are the apsidal motion constants for the host star and planet, respectively, and $c$ is the speed of light \citep{fabrycky_shrinking_2007}. While both inner and outer orbits are subject to relativistic precession, the tidal effects are only applied to the inner planet in our simulations.

\subsection{Evolution during Flybys}
\label{sec:flybys}

When a close encounter within $r_{\rm enc}$ is deemed to occur, the encounter properties are sampled from the appropriate distributions. The components of the field velocity of the perturber $\vec{v}_{\rm pert}$ are first sampled from a Gaussian distribution with the standard deviation corresponding to the local 1D velocity dispersion $\sigma_{\rm v}$. The perturber velocity relative to the host star is then
\begin{equation}
    \vec{v}_{\rm rel} = \vec{v}_{\rm pert} - \vec{v}_{\rm s}
\end{equation}
where $\vec{v}_{\rm s}$ is the velocity of the host star in the cluster and $|\vec{v}_{\rm rel}| = v_\infty$ before gravitational focusing. The primary mass $m_\star$ of the perturber is then drawn from the two-part Kroupa IMF (Equation \ref{eq:imf}) truncated dynamically to exclude stars that have evolved off the main sequence by simulation time $t$:
\begin{equation}
\label{eq:mainsequence}
    m_{\rm \star,\:max}(t) = \left( \frac{t}{10^4~{\rm Myr}} \right)^{-2/5}~{\rm M_\odot}
\end{equation}
based on the standard mass-luminosity relation.\footnote{We note that the truncation of the IMF is only applied in the context of perturber sampling and has no effect on the mean stellar mass $\langle m_\star \rangle$ in the cluster. This is to avoid artificially inflating the stellar number density $n_\star$.} For single-star perturbers, the total mass $m_{\rm pert} = m_\star$. However, a given perturber also has a probability $f_{\rm CB}(m_\star)$ of being a binary. In such cases, the binary mass ratio $q$ is sampled uniformly between 0.1--1 subject to the lower-mass component being in the stellar mass regime (${>} 0.08~{\rm M_\odot}$) such that the total perturber mass becomes $m_{\rm pert} = m_\star (1 + q)$. A uniform $q$ distribution is consistent with observations of dynamically evolved open clusters, including M67 \citep{childs_stellar_2025}. For purposes of numerical efficiency, we restrict the perturber population to close binaries, sampling the binary semi-major axis $a_{\rm bin}$ log uniformly between 1--10~AU and the eccentricity $e_{\rm bin}$ uniformly between 0--1 \citep{moe_mind_2017, geller_stellar_2021}. Based on the total mass and velocity of the perturber, we then calculate the maximum impact parameter $b_{\rm max}$ at which an encounter closer than $r_{\rm enc}$ is possible and finally sample the impact parameter $b$ uniformly in area up to $b_{\rm max}$. The geometry of the encounter $\{ \Omega,\: \omega,\: i\}$ is then sampled isotropically subject to $\vec{b} \cdot \vec{v}_{\rm rel} = 0$.

\subsubsection{Encounter Classification}
\label{sec:flybys_classification}

Of the multiple encounters a planetary system may experience over the cluster lifetime, most are sufficiently distant not to require a direct $N$-body treatment. Here we present a classification scheme for routing encounters into one of three dynamical regimes (analytic, numerical, and hybrid) to allow their impact on the planetary system to be modelled efficiently using a combination of analytic and numerical methods. Specifically, to identify when simpler analytic approximations are valid, we adopt the conditions for a tidal and slow encounter established by \citet{wirth_hot_2025}. For each perturbed body with semi-major axis $a$ and orbital period $P$, we require
\begin{equation}
\label{eq:tidal_criterion}
   \mathcal{T} \coloneq \frac{r_{\rm close}}{a} > \mathcal{T}_{\rm crit}
\end{equation}
and
\begin{equation}
\label{eq:slow_criterion}
    \mathcal{S} \coloneq \frac{\tau_{\rm int}}{P} > \mathcal{S}_{\rm crit}
\end{equation}
where $r_{\rm close}$ is the closest encounter distance of the perturber and the interaction timescale $\tau_{\rm int}$ is estimated as the time interval between the two points along the hyperbolic trajectory where the perturbing force is suppressed by a factor $\xi(r) \equiv (r_{\rm close} / r)^3 = \xi_{\rm crit} = 10^{-4}$. Following \citet{wirth_hot_2025}, we adopt the critical values $\mathcal{T}_{\rm crit} = 15$ and $\mathcal{S}_{\rm crit} = 300$. We note that these threshold values were calibrated for perturbations on a Jovian planet and their applicability to a configuration with a stellar companion (1P1S) was not formally established. However, a limited control experiment in which all flybys were modelled via direct $N$-body integration yielded statistically identical outcomes, confirming that our analytic criteria safely capture all dynamically relevant perturbations.

For binary perturbers, we additionally evaluate the same tidal criterion applied to the intruding binary's own orbit:
\begin{equation}
\label{eq:binary_criterion}
    \mathcal{T}_{\rm bin} \coloneq \frac{r_{\rm close}}{a_{\rm bin}} > \mathcal{T}_{\rm crit}
\end{equation}
which determines whether the internal structure of the perturbing binary is dynamically resolved during the encounter. If $\mathcal{T}_{\rm bin} > \mathcal{T}_{\rm crit}$, the binary is unresolved and the encounter may be treated analytically with the binary perturber replaced by a single point particle of total mass $m_{\rm pert} = m_\star (1 + q)$, provided that the other criteria are satisfied. On the other hand, if $\mathcal{T}_{\rm bin} \leq \mathcal{T}_{\rm crit}$, the binary structure is resolved and the encounter is modelled numerically using direct $N$-body integration regardless of the other criteria. Both components are then integrated explicitly, with their internal orbit determined by the sampled $a_{\rm bin}$ and $e_{\rm bin}$ and initialized with a random orbital phase. When all planetary and stellar companions have been removed from a system, $\mathcal{T}_{\rm bin}$ becomes the sole criterion determining whether an encounter is treated analytically or numerically.

Based on these criteria, each encounter is routed as follows:
\begin{enumerate}
    \item \textbf{Analytic regime} (Section \ref{sec:flybys_analytic}): Both $\mathcal{T}$ and $\mathcal{S}$ criteria (Equations \ref{eq:tidal_criterion}--\ref{eq:slow_criterion}) are satisfied simultaneously for all bound bodies in the system, and any binary perturber is unresolved ($\mathcal{T}_{\rm bin} > \mathcal{T}_{\rm crit}$; Equation \ref{eq:binary_criterion}).
    \item \textbf{Numerical regime} (Section \ref{sec:flybys_numerical}): Either of the $\mathcal{T}$ or $\mathcal{S}$ criteria is violated for the inner planet and therefore for all bound bodies in the system, or a binary perturber is resolved ($\mathcal{T}_{\rm bin} \leq \mathcal{T}_{\rm crit}$).
    \item \textbf{Hybrid regime} (Section \ref{sec:flybys_hybrid}): Both $\mathcal{T}$ and $\mathcal{S}$ criteria are satisfied for the inner planet but not for the outer companion, and any binary perturber is unresolved ($\mathcal{T}_{\rm bin} > \mathcal{T}_{\rm crit}$). This regime combines elements of both approaches described above.
\end{enumerate}

\subsubsection{Analytic Regime}
\label{sec:flybys_analytic}

If all conditions outlined in Section \ref{sec:flybys_classification} are satisfied simultaneously for all bodies in the system, the flyby is in the adiabatic regime in which the semi-major axis remains constant and the secular approximation applies. We can then estimate the eccentricity and inclination excitation $\{\delta e, \delta i\}$ using the analytic formulae derived by \citet{heggie_effect_1996} and \citet{rodet_odea_2019}, respectively (Appendix \ref{sec:formulae}), treating the response of the planet and the companion independently \citep{wang_hot_2022}. While more detailed analytic approximations for binary-binary encounters have been derived \citep{hamers_binarybinary_2020}, these deviate only marginally from the standard binary-single interaction equations. As such, we apply the same analytic approximations also in the case of unresolved binary perturbers which satisfy the tidal criterion ($\mathcal{T}_{\rm bin} > \mathcal{T}_{\rm crit}$), treating the unbound perturber as a single point particle of mass $m_{\rm pert} = m_\star (1 + q)$.

In the analytic regime, the velocity kick on the host star is estimated using the standard formula for hyperbolic deflection and the conservation of momentum. To do so, we first sample a random unit vector $\vec{k}$ perpendicular to the relative velocity $\vec{v}_{\rm rel}$. We then estimate the change in the relative velocity vector based on the exact hyperbolic deflection angle
\begin{equation}
    \tan{\left( \frac{\theta}{2} \right)} = \frac{2 G m_{\rm tot}}{b v_{\rm rel}^2}
\end{equation}
such that the relative velocity after the deflection is given by Rodrigues' rotation formula
\begin{equation}
    \vec{v}_{\rm rel}' = \vec{v}_{\rm rel} \cos{\theta} + (\vec{k} \times \vec{v}_{\rm rel}) \sin{\theta} + \vec{k}(\vec{k} \cdot \vec{v}_{\rm rel}) (1 - \cos{\theta})
\end{equation}
where the last term vanishes by definition. Using the change in the relative velocity $\Delta \vec{v}_{\rm rel} = \vec{v}_{\rm rel}' - \vec{v}_{\rm rel}$, the corresponding change in the host velocity in the cluster frame is then
\begin{equation}
    \Delta \vec{v}_{\rm s} = -\frac{m_{\rm pert}}{m_{\rm tot}} \Delta \vec{v}_{\rm rel}
\end{equation}
which is propagated to the \texttt{gala} simulation to self-consistently model the stellar trajectory in the cluster. We avoid double-counting velocity kicks from weak encounters by explicitly evaluating the Coulomb logarithm for stochastic diffusion as $\ln{\Lambda} = \ln{(r_{\rm t} / r_{\rm enc})}$ where the radius of the encounter sphere ($r_{\rm enc}$) sets the minimum impact parameter for stochastic diffusion due to unresolved background perturbations (Section \ref{sec:stellardynamics}; \citealt{chandrasekhar_dynamical_1943}).

\subsubsection{Numerical Regime}
\label{sec:flybys_numerical}

Any encounters not satisfying both tidal ($\mathcal{T}$) and slow ($\mathcal{S}$) criteria for all bound bodies, as well as resolved binary encounters ($\mathcal{T}_{\rm bin} \leq \mathcal{T}_{\rm crit}$), are simulated directly using the adaptive \texttt{IAS15} integrator in \texttt{REBOUND} \citep{rein_ias15_2015}. In such cases, the perturber is initialized at the boundary of the encounter sphere at a distance $r_{\rm enc}$ defined in Equation \ref{eq:renc}. The integration then proceeds until the perturber recedes to a distance $r > r_{\rm enc}$ following the close encounter. We tested the validity of this initialization boundary by experimenting with larger starting distances in line with the critical radii used for determining the adiabatic interaction timescale. These experiments confirmed that extending the numerical integration beyond $r_{\rm enc}$ yields negligible differences in the final orbital elements while incurring a significantly higher computational cost, consistent with the rapid fall-off of the tidal force with distance ($F_{\rm tide} \propto r^{-3}$).

Following the flyby, changes in the orbital elements of the planet and its potential companion relative to the host star are recorded accordingly. Rather than estimating the velocity kick on the host star using the analytic formulae, which may break down in the case of close encounters with binary perturbers, $\Delta \vec{v}_{\rm s}$ is evaluated directly based on the change in the centre-of-mass velocity vector of the planetary system during the flyby in \texttt{REBOUND} and propagated to the \texttt{gala} simulation. To ensure self-consistency with the cluster simulation, if any component of the system is ejected as a result of the close encounter, the final centre-of-mass velocity is computed based only on the remaining bound bodies.

\subsubsection{Hybrid Regime}
\label{sec:flybys_hybrid}

In systems with a significant separation of scales between the inner and outer orbits, direct $N$-body integration of the full system during a flyby can become computationally prohibitive. Since an adaptive integrator such as \texttt{IAS15} scales its internal timestep based on the shortest dynamical timescale present in the system, a hierarchical configuration with an inner planet on a short-period orbit and an outer companion on wide orbit forces the integrator to take a large number of tiny steps to resolve the inner orbit, even while the perturber is interacting almost exclusively with the outer companion. Given that the tidal force scales as $F_{\rm tide} \propto r^{-3}$, a perturber interacting primarily with the outer companion at $a_{\rm c}$ exerts a negligible direct tidal force on the inner planet at $a_{\rm p} \ll a_{\rm c}$. Full numerical integration of the inner orbit during such an encounter is therefore both computationally costly and physically unnecessary.

In the hybrid regime, the inner planet is excluded from the \texttt{REBOUND} simulation and its orbital response to the flyby is calculated analytically using the \citet{heggie_effect_1996} and \citet{rodet_odea_2019} formulae (Appendix \ref{sec:formulae}). The numerical integration then proceeds with only the host star, outer companion, and unbound perturber, allowing the integrator to take timesteps governed by the outer orbital period rather than the inner one. Following the flyby, the velocity kick on the host star is extracted directly from the \texttt{REBOUND} simulation, similarly to the numerical regime (Section \ref{sec:flybys_numerical}), and the mean anomaly of the inner planet is advanced analytically based on its mean motion and the elapsed integration time during the flyby. We note that the exclusion of the inner planet from the \texttt{REBOUND} simulation marginally affects the centre-of-mass velocity from which $\Delta \vec{v}_{\rm s}$ is extracted. However, since $m_{\rm p} \ll m_{\rm s}$, the planet contributes negligibly to the total mass of the system and this effect is safely ignored. We verified via a control experiment that full $N$-body integration of encounters routed to this regime yields statistically indistinguishable outcome fractions, while the hybrid approach results in a substantial speedup in the overall runtime compared to the numerical regime.

\subsection{Outcome Classification}
\label{sec:outcome_classification}

To describe the final state of each system, we define a set of classifications similar to those used by \citet{hamers_hot_2017} and \citet{wirth_hot_2025}:
\begin{enumerate}
    \item \textbf{Ionization (I)}: A planet is considered unbound from its host star and therefore ionized if its eccentricity exceeds unity ($e \geq 1$) or its apoapsis distance exceeds the Hill radius of the system:
    \begin{equation}
        r_{\rm apo} = a(1 + e) \geq r_{\rm Hill} \equiv r_{\rm s} \times \left( \frac{m_{\rm s}}{M_\star({<} r_{\rm s})} \right)^{1/3}
    \end{equation}
    where $r_{\rm s}$ is the cluster-centric radius of the host star and $M_\star({<} r_{\rm s})$ is the enclosed cluster mass. The companion is also considered as ionized if its orbital period exceeds the simulation timestep ($\delta t = 0.1~{\rm Myr}$), corresponding to $a \gtrsim 2,000~{\rm AU}$.\footnote{This threshold exceeds the maximum hard-soft boundary reached for equal-mass binaries over the course of the simulation by a factor of $\sim$2, implying that systems wider than this are prone to rapid dissociation by the cluster field.} In the event of a simple orbit crossing ($a_{\rm p} > a_{\rm c}$), the more weakly bound of the two bodies is considered to be ionized.
    \item \textbf{Tidal Disruption (TD)}: A planet is considered tidally destroyed if its periapsis distance is closer than the tidal disruption radius:
    \begin{equation}
    \label{eq:td_limit}
        r_{\rm peri} = a(1 - e) < \eta r_{\rm Roche} \approx \eta R_{\rm p} \left( \frac{m_{\rm s}} {m_{\rm p}} \right)^{1/3}
    \end{equation}
    where we set $\eta = 2.7$ based on numerical results by \citet{guillochon_consequences_2011}.
    \item \textbf{Hot Jupiter (HJ)}: A planet is considered to be a hot Jupiter if its semi-major axis $a < 0.1$~AU.
    \item \textbf{Hot Jupiter Candidate (HJC)}: A planet is considered to be a hot Jupiter candidate if its periapsis distance $r_{\rm peri} < 0.1$~AU.
    \item \textbf{Warm Jupiter (WJ)}: A planet is considered to be a warm Jupiter if its semi-major axis is in the range $ 0.1~{\rm AU} \leq a < 1~{\rm AU}$ and it is not a HJC.
    \item \textbf{Cold Jupiter (CJ)}: A planet is considered to be a cold Jupiter if its semi-major axis $a \geq 1$~AU and it is not a HJC.
\end{enumerate}

If the orbit of a planet shrinks sufficiently ($a < 0.1$~AU) for it to be classified as a HJ, we deem it decoupled from its potential outer companion as well as the wider stellar environment and assume its subsequent evolution until the end of the simulation is dominated by tides. At such short orbital periods, GR and tidal precession suppress ZLK oscillations on timescales shorter than the companion's orbital period, effectively quenching any further eccentricity excitation and allowing migration to proceed independently of the outer orbital architecture \citep{fabrycky_shrinking_2007}. Similarly, if either of the tidal disruption or ionization conditions is satisfied at any point, the corresponding body is removed from the active simulation.

The balance between HJ formation and tidal disruption is sensitive to several poorly constrained aspects of the tidal model. In particular,  \citet{teyssandier_formation_2019} found the choice of the tidal disruption radius -- commonly set to 2.7~$r_{\rm Roche}$ following \citet{guillochon_consequences_2011} -- to significantly affect the number of surviving HJs. The inclusion of chaotic dynamical tides \citep{vick_chaotic_2019} and a realistic prescription for partial mass loss \citep{weldon_saving_2026} may boost the fraction of tidally disrupted systems that survive as hot Jupiters. Given these uncertainties, we also consider an alternative outcome classification in which the HJ, HJC, and TD stopping conditions are merged to reflect a more optimistic hot Jupiter yield. We discuss the implications of these effects for the inferred occurrence rate further in Section \ref{sec:occurrencerate}.

\subsection{Suite of Simulations}

To isolate the specific impact of environmental perturbations on planetary architectures, we perform two distinct sets of simulations for each of the three planetary configurations described in Section \ref{sec:planetarysystem}. In addition to the M67-like cluster scenario described above, we perform a control study in the field with no flybys ($\Gamma = 0$). In other aspects, the simulation is identical. This control sample accounts for any intrinsic dynamical instabilities arising purely from the initial orbital configurations (e.g. planet-planet scattering or ZLK oscillations in primordial binaries) over the $\sim$4~Gyr integration time.

For each configuration (single planet, planet + planet, planet + star), we simulate 10,000 realizations in both the cluster and field scenarios. Given our sampling of initial planetary system parameters and stochastic treatment of stellar flybys, this ensures adequate coverage of the full parameter space. Since the RV survey by \citet{brucalassi_search_2016} specifically targeted member stars in the core of M67, we split the sample of bound cluster stars in two based on whether their final cluster-centric radius is within the half-mass radius (3.8~pc) at the end of the simulation. Planetary systems which escape the cluster are also considered as a separate sample.

\section{Results}
\label{sec:results}

\subsection{Cluster vs Field Sample}
\label{sec:clusterfield}

\begin{table*}
\centering
\begin{tabular}{ll|ccccccc}
\hline
& & \multicolumn{6}{c}{Outcome Fraction (per cent)} & \\
Configuration & Sample & I & TD & HJ & HJC & WJ & CJ & $N$ \\
\hline
\multirow{5}{*}{1P}
 & Cluster               & 0.69(9) & 0 & 0.01(1) & 0.01(1) & 0.01(1) & 99.27(9) & 8,128 \\
 & -- $r < r_{\rm h}$    & 0.7(1) & 0 & 0.02(2) & 0.02(2) & 0.02(2) & 99.3(1) & 6,561 \\
 & -- $r \geq r_{\rm h}$ & 0.7(2) & 0 & 0 & 0 & 0 & 99.3(2) & 1,567 \\
 & Escapers              & 0.6(2) & 0 & 0 & 0 & 0.11(8) & 99.3(2) & 1,872 \\
 & Field                 & 0 & 0 & 0 & 0 & 0 & 100 & 10,000 \\
\hline
\multirow{5}{*}{2P}
 & Cluster               & 1.8(1) & 1.5(1) & 0.12(4) & 0.04(2) & 0.02(2) & 96.5(2) & 8,023 \\
 & -- $r < r_{\rm h}$    & 2.0(2) & 1.6(2) & 0.09(4) & 0.05(3) & 0.02(2) & 96.3(2) & 6,409 \\
 & -- $r \geq r_{\rm h}$ & 1.2(3) & 1.0(2) & 0.2(1) & 0 & 0.06(6) & 97.5(4) & 1,614 \\
 & Escapers              & 1.2(2) & 0.6(2) & 0 & 0 & 0 & 98.2(3) & 1,977 \\
 & Field                 & 0 & 0 & 0 & 0 & 0 & 100 & 10,000 \\
\hline
\multirow{5}{*}{1P1S}
 & Cluster               & 10.4(3) & 15.0(4) & 10.8(3) & 0.72(9) & 0.23(5) & 62.9(5) & 8,222 \\
 & -- $r < r_{\rm h}$    & 11.1(4) & 15.7(4) & 11.1(4) & 0.7(1) & 0.24(6) & 61.2(6) & 6,957 \\
 & -- $r \geq r_{\rm h}$ & 6.6(7) & 11.3(9) & 9.2(8) & 0.9(3) & 0.2(1) & 72(1) & 1,265 \\
 & Escapers              & 3.9(5) & 6.6(6) & 7.0(6) & 0.5(2) & 0.2(1) & 81.9(9) & 1,778 \\
 & Field                 & 0 & 4.6(2) & 5.2(2) & 0.45(7) & 0.05(2) & 89.7(3) & 10,000 \\
\hline
\end{tabular}
\caption{Outcome fractions for the inner planet across the three initial configurations (1P, 2P, 1P1S) in the cluster and field samples. Each row corresponds to a distinct subsample: bound cluster stars are further divided based on whether their final cluster-centric radius is within ($r < r_{\rm h}$) or beyond ($r \geq r_{\rm h}$) the half-mass radius, and stars that escape the cluster are listed separately. Each percentage indicates the fraction of systems within that subsample whose inner planet attained the corresponding dynamical outcome by the end of the 4~Gyr integration. We also quote the associated uncertainties evaluated as binomial standard errors $\sigma = \sqrt{f(1-f)/N}$ rounded to one significant figure in parentheses. Note that the fractions in some rows do not sum to exactly 100~per cent due to rounding.}
\label{table:outcomes}
\end{table*}

Table \ref{table:outcomes} summarizes the fraction of systems in which the inner planet attains a given dynamical state across the three different initial configurations. In the 1P case of a single Jovian planet initialized at a semi-major axis between 1--10~AU, we find the cluster environment to have limited effect on the fate of the system. While flybys can induce high eccentricities (leading to a $\sim$0.7~per cent ionization rate), they rarely trigger the precise, high-eccentricity conditions required for tidal migration. Overall, only one hot Jupiter is formed through direct scattering by flybys in the densest part of the cluster. Additionally, planets in three 1P systems undergo sufficient inward migration to end up as warm Jupiters, with two such systems ultimately escaping the cluster. Given that they evolve solely under the influence of tides that are only activated for $r_{\rm peri} < 0.5$~AU, single-planet systems in the field sample remain unchanged from their initial conditions.

Similarly, in the 2P configuration, the HJ yield remains low at $\sim$0.1~per cent in the cluster. While both ionization and tidal disruption occur at a rate of 1--2~per cent, only ten hot Jupiters are formed in the entire simulation run. Of these, six retain a bound outer planetary companion until the end of the integration, the remainder having been ionized. This suggests that while planet-planet scattering can be triggered by external perturbations, it is an inefficient pathway for generating stable hot Jupiters in a relatively low-density open cluster environment. Instead, violent instabilities are more likely to result in tidal disruption or ejection than tidal capture. In the isolated field simulation, all two-planet systems retain their original configurations, as expected from our choice of stable initial conditions. It is possible that the choice of broader eccentricity and inclination distributions would lead to some intrinsic HJ formation in the field 2P sample. In particular, previous works on the evolution of single-planet systems in clusters \citep{hamers_hot_2017, wirth_hot_2025} have adopted a Rayleigh eccentricity distribution with scale parameter 0.33 reflecting the effect of post-disc-phase scattering \citep{juric_dynamical_2008, ford_origins_2008}. However, given the small number of significantly modified systems in our sample, such a choice would complicate the interpretation of the cluster enhancement factor. We additionally note that, even when the stability and hierarchy criteria outlined in Section \ref{sec:planetarydynamics} are satisfied, weak interactions between the planetary orbits may drive angular momentum and energy exchange not captured by the orbit-averaged secular approximation, and the 2P hot Jupiter yield should therefore be regarded as a lower limit. However, to explicitly assess their stability, we performed direct $N$-body integrations of 100 isolated two-planet systems for 10~Myr, which yielded no significant deviations from the initial configurations, demonstrating that these architectures are intrinsically quiescent in the absence of external perturbations.

The most significant environmental impact is observed in the 1P1S scenario. In the isolated field sample, the HJ and TD yields are 5.2 and 4.6~per cent, respectively, driven entirely by primordial high-inclination configurations undergoing ZLK oscillations. Notably, in the cluster core, the HJ formation rate is doubled to 11.1~per cent, while the TD rate is more than tripled to 15.7~per cent. Both yields decrease towards the lower-density outskirts of the cluster, but a slight increase in both the HJ and TD rates compared to the field is still evident among the escaping population, an aspect that we discuss further in Section \ref{sec:cluster_evolution}. Considering all systems with final a periapsis distance below 0.1~AU, the cluster-induced excess appears even more pronounced, with the combined HJ + HJC + TD outcome yield standing at 10.3~per cent in the field compared to 27.5~per cent in the cluster core. This demonstrates the role of the cluster environment in destabilizing initially quiescent binary systems and driving them into configurations where strong eccentricity pumping of the inner planet is possible. Additionally, 10.4~per cent of inner planets in the cluster sample undergo complete ejection (compared to 0.0~per cent in the field), establishing cluster-induced destabilization as an efficient dynamical channel for generating free-floating planets (FFPs). Given the strong enhancement in HJ formation rates in the 1P1S case, we will primarily focus on this configuration in our analysis.

\subsection{Role of Initial Conditions}

\begin{figure*}
\includegraphics[width=\textwidth]{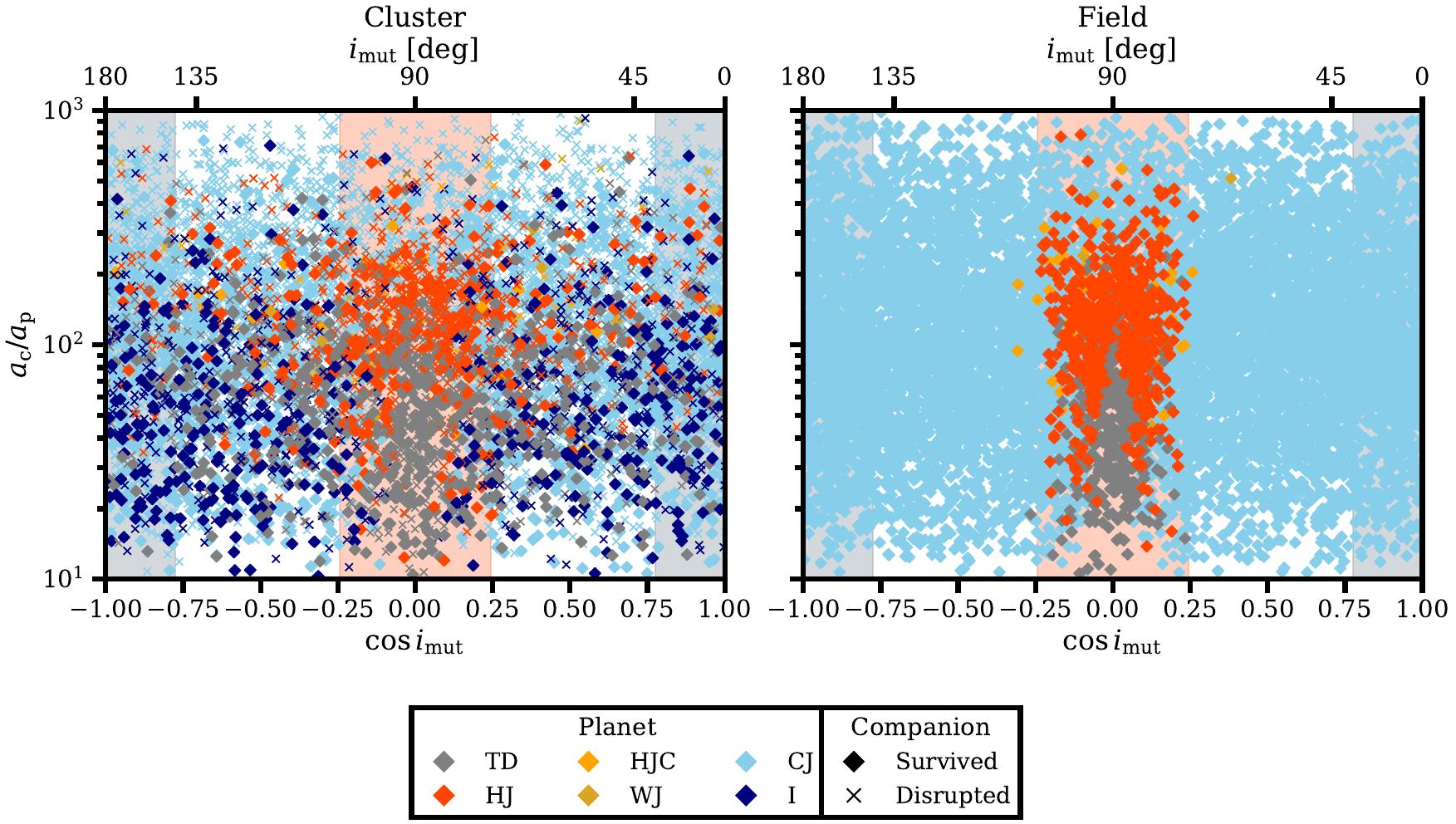}
\caption{Outcomes for 1P1S-type planetary systems as a function of the initial mutual inclination angle (horizontal) and the ratio of the outer and inner semi-major axes used in the \citet{mardling_tidal_2001} stability criterion (Equation \ref{eq:mardling}). The two panels correspond to the cluster (left) and field (right) samples. The grey shaded regions designate the range of inclination angles at which significant ZLK oscillations cannot occur ($e_{\rm max}$ is undefined), whereas systems in the orange shaded region can drive a planet initially at 1~AU to a sufficiently high eccentricity ($e_{\rm p} \gtrsim 0.95$) for HJ formation via tidal migration based on initial conditions alone (see text).
\label{fig:smaratio_cosimut}}
\end{figure*}

\begin{figure*}
\includegraphics[width=0.8\textwidth]{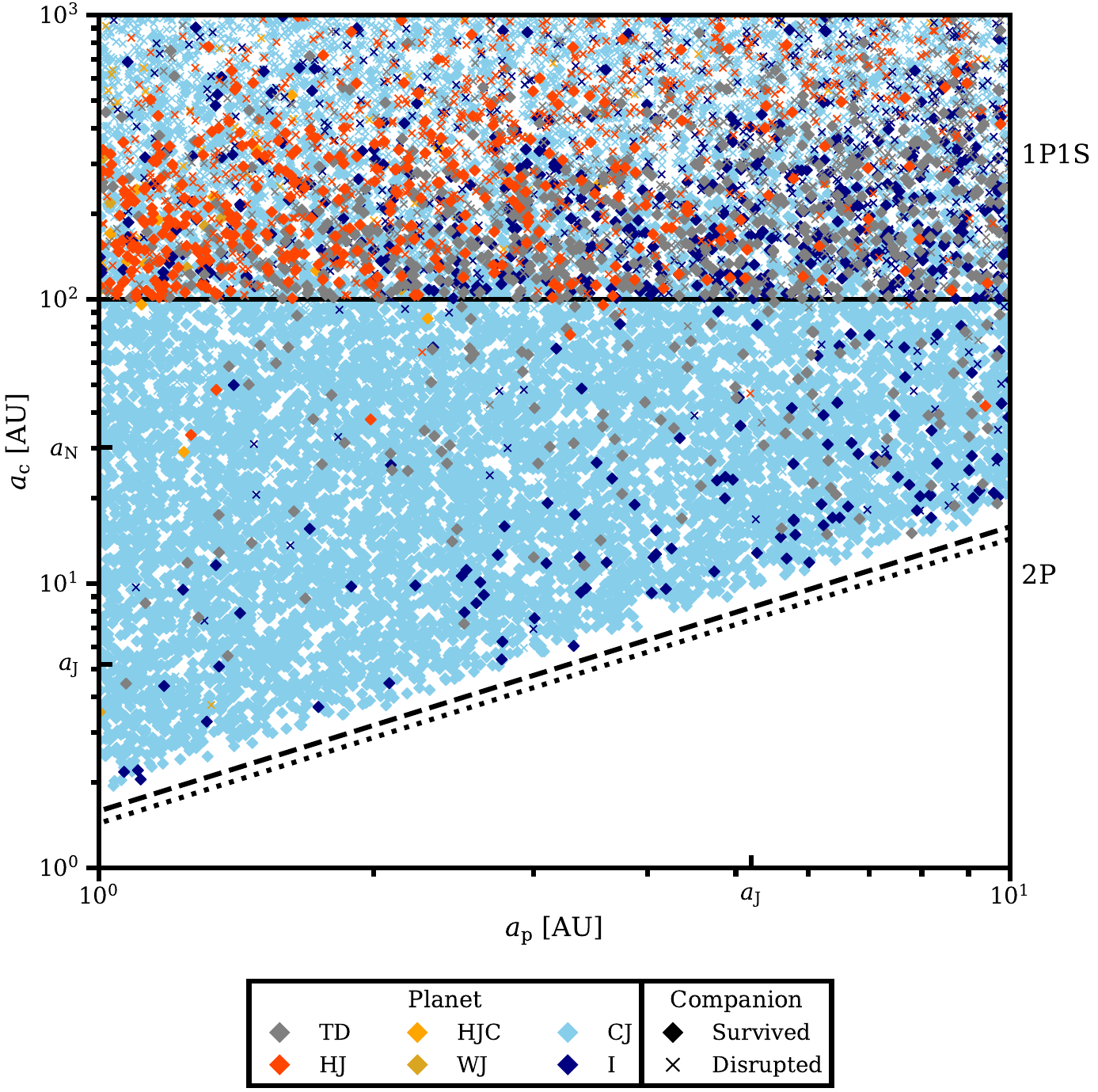}
\caption{Outcomes for planetary systems retained in the cluster as a function of the initial semi-major axes of the inner planet (horizontal) and outer companion (vertical). The solid horizontal line separates systems initialized with a planetary (2P, below) and a stellar (1P1S, above) companion. The black dashed line corresponds to the 2:1 mean-motion resonance and the black dotted line to the \citet{petrovich_stability_2015} stability criterion assuming circular, coplanar orbits and a 1~$\rm M_J$ outer planetary companion. The semi-major axes of Jupiter ($a_{\rm J}$) and Neptune ($a_{\rm N}$) are shown for reference.
\label{fig:ain_aout}}
\end{figure*}

Insight into the mechanisms driving the enhanced rate of hot Jupiter formation and tidal disruption events in the cluster environment may be gained from considering the initial conditions of these systems. As shown in Figure \ref{fig:smaratio_cosimut} for the 1P1S configuration, the formation of hot Jupiter candidates in the control sample is strictly predicated on a specific set of initial conditions, particularly an initial mutual inclination angle close to 90$^\circ$. In contrast, HJs and HJ candidates form across a wide range of initial inclination angles in the cluster environment. This reflects the tendency of both cumulative weak encounters and strong flybys to torque the outer stellar companion, effectively randomizing the mutual inclination plane. Systems with initially low mutual inclinations are therefore pumped into the ZLK window \citep{winter-granic_binary_2024}, facilitating HJ formation. The timeframe for HJ formation is correspondingly extended, with the median formation time in the cluster standing at $\sim$370~Myr, compared to just $\sim$60~Myr in the field, reflecting the additional time required to torque initially quiescent systems into the ZLK window before high-eccentricity migration can proceed. 

Also evident in Figure \ref{fig:smaratio_cosimut} is the role of the initial dynamical spacing in the 1P1S configuration, quantified here using the semi-major axis ratio of the outer and inner orbits, also used in the \citet{mardling_tidal_2001} stability criterion. At small values ($a_{\rm c}/a_{\rm p} \lesssim 50$), the growing octupole perturbations drive eccentricities to near-unity on a timescale shorter than the tidal response time, leading the inner planet to undergo prompt disruption rather than being gradually circularized. On the other hand, at large values ($a_{\rm c}/a_{\rm p} \gtrsim 500$), the two orbits become dynamically decoupled, with the torque on the inner orbit becoming too weak to drive efficient eccentricity excitation. This may partly reflect quenching of ZLK oscillations by general relativistic precession which suppresses eccentricity growth when $\dot\omega_{\rm GR} > \dot\omega_{\rm ZLK}$ \citep{liu_suppression_2015}. This condition is most readily satisfied for tight inner orbits and wide outer companions, consistent with the absence of HJ formation at large $a_{\rm c}/a_{\rm p}$ in the field sample. Between these lies a sweet spot for HJ formation via HEM.

Furthermore, the initial semi-major axes of the system components play a crucial role in determining the final outcome in both the cluster and field samples, as shown in Figure \ref{fig:ain_aout}. The upper part of the parameter space, corresponding to the 1P1S configuration, suggests the presence of a band of semi-major axis ratios along which HJ formation is favoured, as demonstrated in Figure \ref{fig:smaratio_cosimut}. Hot Jupiter formation is particularly prevalent among systems with an initial planetary semi-major axis below $\sim$3~AU. Beyond this distance, the requisite eccentricity for circularization ($e \gtrsim 0.98$) brings the planet dangerously close to the tidal disruption radius (Equation \ref{eq:td_limit}) or facilitates orbit crossing with the companion, leading to ionization.

The lower part of the parameter space corresponds to the 2P case and suggests more quiescent dynamical evolution, with most systems still consisting of two cold Jupiters at the end of the simulation. This is expected given our choice of relatively coplanar, low-eccentricity initial conditions. In particular, inner planet ionization occurs mainly among systems that are initialized beyond $\sim$3~AU and are close to the \citet{petrovich_stability_2015} stability boundary. The closer dynamical spacing makes strong interactions between the orbits more common, resulting in an increased rate of ionization. In our limited sample, HJ and HJ candidate (HJC) formation only occurs in systems with the outer planet initially beyond 30~AU, as only such wide companions are efficiently kicked into the ZLK window. Notably, HJ formation in 2P systems occurs across a wide range of inner planet orbital radii. In contrast, both tidal disruption and ionization due to planet-planet scattering preferably occur at large initial values of $a_{\rm p}$.

In the 1P1S case, we find a slight positive correlation between the mass of the stellar companion and HJ formation rate in the cluster sample. The strength of this correlation reflects a balance between two competing effects: while more massive companions drive faster angular momentum transfer through their effect on the quadrupole timescale $\tau_{\rm ZLK}$ (Equation \ref{eq:quadrupole}), favouring HJ formation, more massive systems also experience more frequent encounters due to gravitational focusing, thereby increasing the rate at which systems diffuse both into and out of the ZLK window, partially offsetting this trend. No clear correlation arises in the control sample, consistent with HJ formation in the field not being significantly limited by timescale constraints given that our integration timescale (4~Gyr) exceeds the longest quadrupole timescale in our sample ($\sim$960~Myr). On the other hand, the tidal disruption fraction increases monotonically with companion mass in both cluster and field samples, consistent with more massive companions driving faster and more extreme ZLK cycles that push the inner planet's eccentricity to values where tidal disruption rather than circularization becomes the more probable outcome.

The outcome fractions in our control sample with a stellar companion can be interpreted through simple analytic estimates. At the quadrupole order, the maximum eccentricity that can theoretically be attained by the inner orbit during a ZLK cycle is given by
\begin{equation}
    e_{\rm max} = \sqrt{1 - \frac{5}{3} \cos^2{i_{\rm mut}}}
\end{equation}
where $i_{\rm mut}$ is the mutual inclination between the orbits. Based on the initial distribution of orbital elements, we can therefore estimate the fraction of systems that are initially susceptible to HJ formation. For the sake of simplicity, we assume the angular momentum $a(1-e^2) \sim {\rm constant }$ during the circularization. In order to tidally circularize with a final semi-major axis of 0.1~AU, a planet initially at 1~AU must therefore attain an eccentricity of approximately 0.95, corresponding to a periapsis distance of $\sim$0.05~AU. The fraction of systems with initial attainable periapsis distances below this value yet above the tidal disruption limit (Equation \ref{eq:td_limit}) is $\sim$5.0~per cent, whereas $\sim$4.8~per cent of systems are expected to be disrupted based on the initial conditions alone. These analytic estimates are in fair agreement with the outcome fractions from our control simulation without flybys. The slight differences between the predicted and realized outcome fractions are explained by octupole-order effects and our inclusion of dynamical tides with $a(1-e) \sim {\rm constant }$ in the model. We show the range of mutual inclination angles corresponding to the circularization of a planet initially at 1~AU at a final orbital radius of 0.1~AU as an orange shaded region in Figure \ref{fig:smaratio_cosimut}.

\subsection{Companion Survival}

\begin{figure*}
\includegraphics[width=0.8\textwidth]{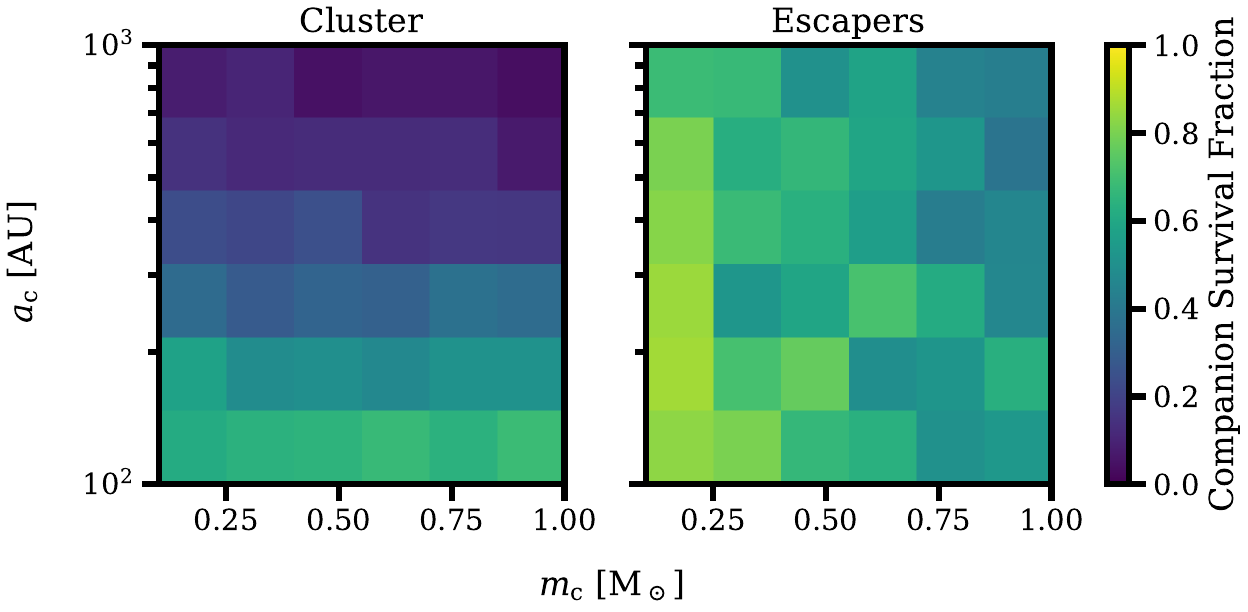}
\caption{The fraction of surviving outer stellar companions as a function of mass ($m_{\rm c}$) and initial semi-major axis ($a_{\rm c}$) in the bound (left) and escaping (right) populations. While the cluster environment efficiently ionizes wider binaries ($a_{\rm c} \gtrsim 300~{\rm AU}$) regardless of companion mass, a slight mass dependence emerges among the escaping population due to more massive systems remaining longer in the cluster before escaping and therefore being more susceptible to disruption.
\label{fig:companion_survival}}
\end{figure*}

In the cluster environment, few wide stellar companions survive the full 4~Gyr integration, with the survival fraction declining below $\sim$20~per cent for initial semi-major axes $a_{\rm c} \gtrsim 300~{\rm AU}$, as shown in Figure \ref{fig:companion_survival}. Among the bound cluster members, systems that successfully form hot Jupiters exhibit an elevated companion survival rate ($36 \pm 2$~per cent) compared to those in which the planet remains a cold Jupiter ($23 \pm 1$~per cent). This reflects a dynamical selection bias due to HJ formation heavily favouring tighter ($a_{\rm c} \lesssim 300~{\rm AU}$) and more massive stellar companions capable of driving rapid ZLK cycles. These tighter configurations are more resilient to ionization, which elevates their survival rate in the dense core relative to the broader CJ population. Nevertheless, with $\sim$2/3 of HJ-hosting systems having lost their companions, the lack of observed stellar companions to the three hot Jupiter hosts in M67 \citep{brucalassi_search_2016} cannot be taken as strict evidence against their formation via binary-driven high-eccentricity migration.

For comparison, in the 2P configuration the outer planetary companion is retained in $60 \pm 20$~per cent of HJ-hosting cluster systems, significantly lower than the $92.1 \pm 0.3$~per cent survival rate among systems in which the inner planet remains a cold Jupiter. While this comparison is based on a small number ($N = 10$) of HJ-forming two-planet systems, it is notable that the trend is reversed relative to the 1P1S configuration. This reflects the different migration mechanism acting in the two configurations: unlike the comparatively gentle ZLK migration underlying the 1P1S case, HJ formation via planet-planet scattering is itself a disruptive process that frequently ejects the outer companion, rather than preferentially selecting for companions that are inherently more resilient to disruption. The survival fraction in our simulations is broadly consistent with the observed rate of $70 \pm 8$~per cent of hot Jupiter systems hosting an outer planetary companion \citep{knutson_friends_2014, bryan_statistics_2016}.

Among the escaping population, stellar companion survival is largely independent of the dynamical outcome of the inner planet, with $64 \pm 4$~per cent of HJ hosts and $64 \pm 1$~per cent of CJ hosts retaining their stellar companions. These values are somewhat higher (at the ${\sim}2\sigma$ level) than the observed fraction ($47 \pm 7$~per cent) of HJ-hosting systems with stellar companions in the range 50–2,000 AU \citep{ngo_friends_2016}. However, since flybys are switched off once a system crosses the cluster's tidal boundary (Section \ref{sec:stellardynamics}), our model does not capture any subsequent companion disruption via the Galactic tidal field or field star encounters over the remaining time since escape. We therefore regard our escaper survival fractions as upper limits on the true present-day incidence of stellar companions among long-escaped HJ hosts, which may partially account for the slight tension with the \citet{ngo_friends_2016} measurement. While the distribution of $a_{\rm c}$ among surviving systems is shifted towards slightly larger values, it remains statistically indistinguishable from the initial log uniform distribution used for sampling and is in broad agreement with observations in the field.

The overall high companion survival fraction among the escaping population suggests that most of these systems leave the cluster via gentle tidal stripping by the Galactic potential, preserving their primordial architectures. This is further supported by the low median velocity (${\sim} 0.6~{\rm km~s^{-1}}$) of escaping stars at the point of crossing the tidal boundary. As shown in Figure \ref{fig:companion_survival}, a slight mass dependence is discernible at intermediate semi-major axes in the escaping sample, with lower-mass companions surviving preferentially. This reflects the fact that more massive systems spend more time in the cluster before escaping, accumulating a larger number of encounters and therefore facing a higher cumulative probability of companion ionization. In the cluster, by contrast, all systems are exposed to the collisional environment for the full $\sim$4~Gyr integration, and the weak dependence of survival on companion mass instead reflects a trade-off between two competing effects: while lower-mass companions are energetically easier to ionize, more massive companions experience more frequent close encounters due to gravitational focusing.

Particularly destructive are penetrating flybys, defined here as those with a closest encounter distance $r_{\rm close}$ smaller than the initial outer semi-major axis $a_{\rm c}$. In our simulation, up to 82~per cent of all 1P1S systems experience one such flyby over the 4~Gyr integration, consistent with the expectation from the focused rate (Equation \ref{eq:gamma_focused}) evaluated at $r_{\rm enc} = a_{\rm c}$. Of these, 75~per cent end up losing their stellar companions. By contrast, only 13~per cent of systems that do not experience a penetrating flyby lose their companions. Close flybys also have dramatic implications for the survival of the inner planet, with the corresponding inner planet ionization fractions being 11~per cent (3~per cent) for systems with (without) penetrating flybys. On the other hand, penetrating encounters are only responsible for a small fraction of planetary systems escaping the cluster. Only 10~per cent of such systems escape the cluster, compared to 46~per cent of those that only ever experienced more distant encounters. This reflects yet another dynamical selection bias: only systems that remain bound to the cluster are likely to experience a very close encounter. Conversely, hosts that either originate in or rapidly migrate to the outskirts of the cluster are not only less likely to ever experience a penetrating encounter but also more likely to escape through tidal stripping and preserve their primordial architectures.

\subsection{Dynamical Sculpting of Surviving Systems}
\label{sec:dynamicalsculpting}

\begin{figure*}
\includegraphics[width=\textwidth]{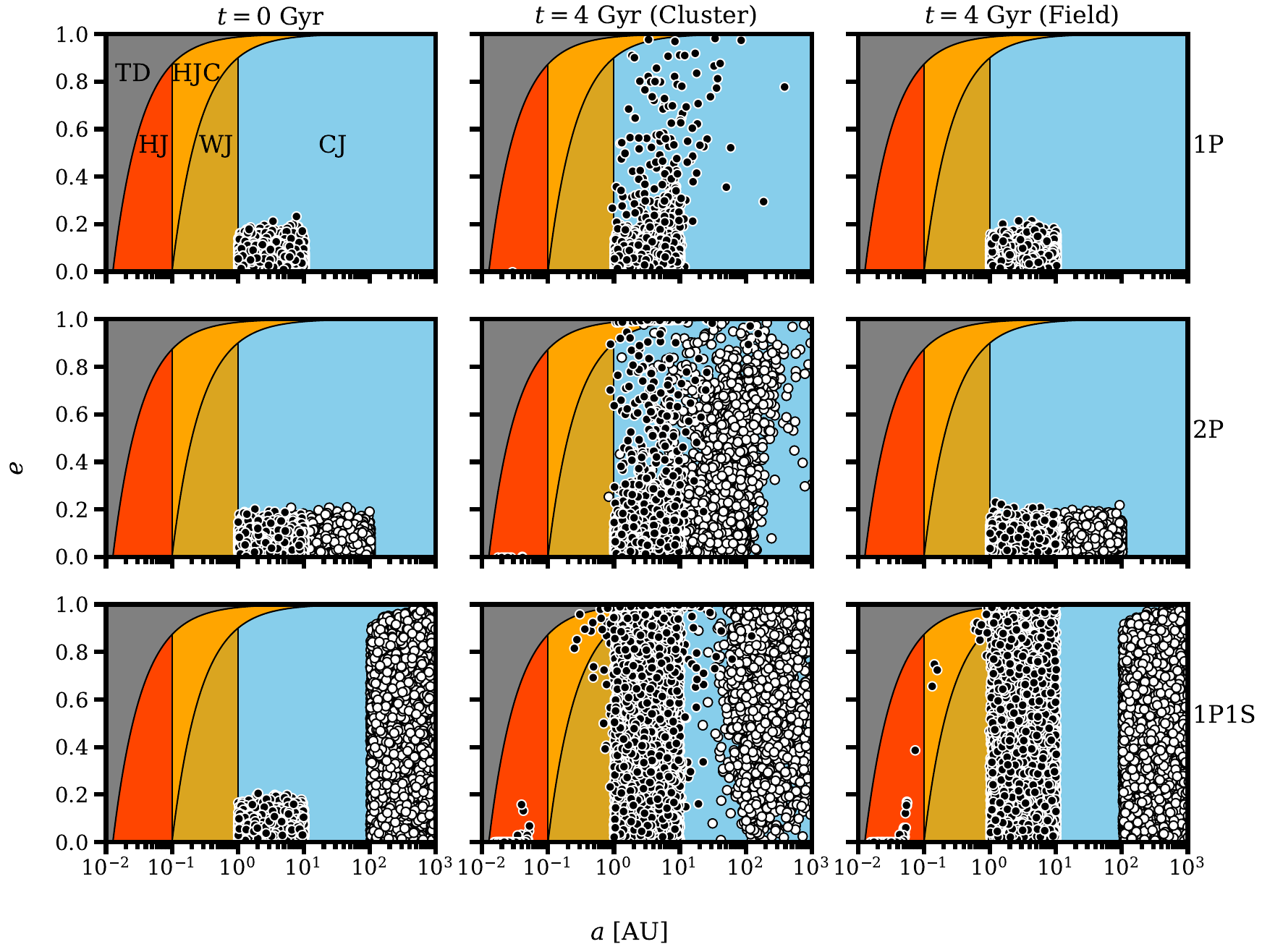}
\caption{The eccentricity and semi-major axis distribution of the planet (black) and companion (white) at the beginning (left) and the end of the simulation with (middle) and without (right) flybys across three sets of initial conditions (1P, 2P, 1P1S). While only the initial conditions for the cluster sample are included in the left-hand column, those for the field control sample are statistically indistinguishable. The parameter space is divided based on the outcome classification of the inner planet (Section \ref{sec:outcome_classification}), with annotations provided in the top left panel, and the colour scheme matches that used in Figures \ref{fig:smaratio_cosimut} and \ref{fig:ain_aout}.
\label{fig:sma_ecc}}
\end{figure*}

\begin{figure*}
\includegraphics[width=0.9\textwidth]{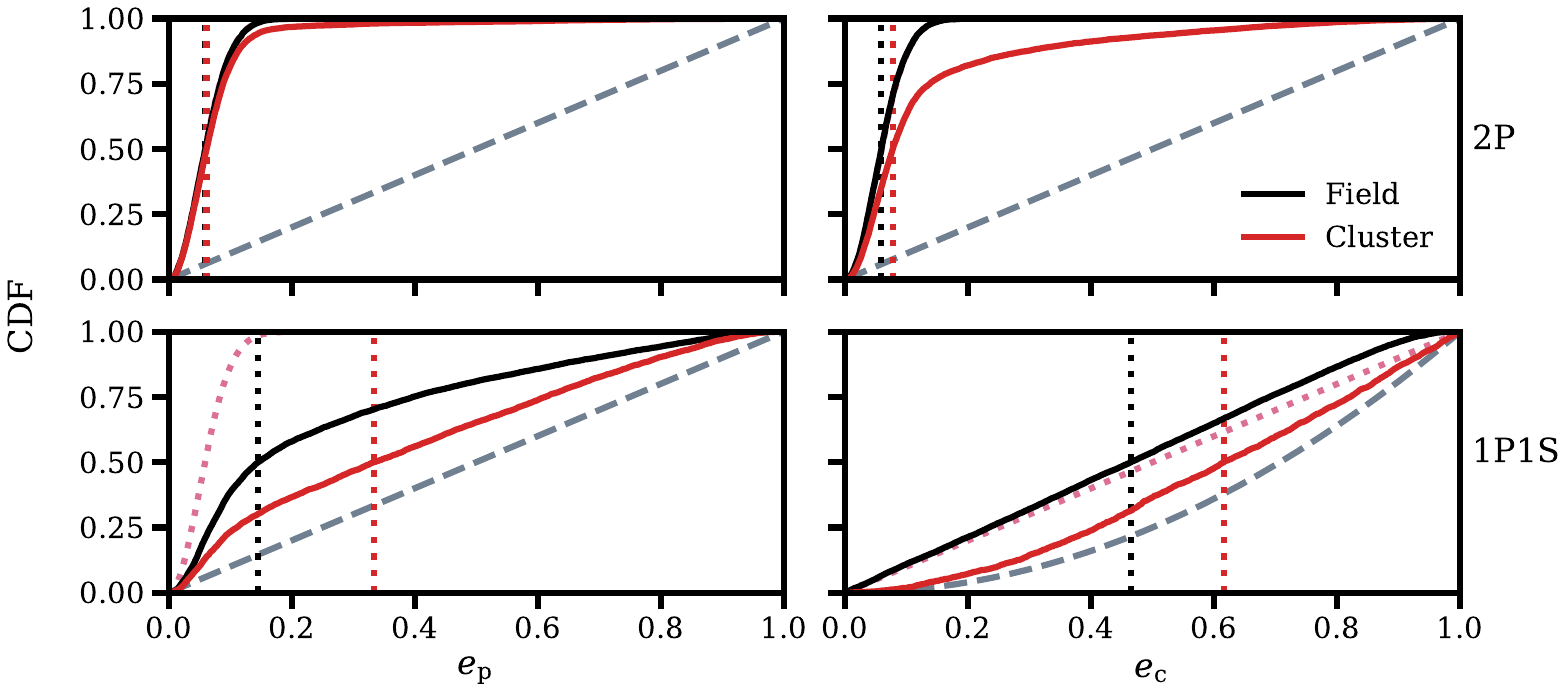}
\caption{Final eccentricity distributions of inner planets remaining as cold Jupiters (left) and surviving outer companions (right) in the cluster (red) and field (black) samples for the 2P (top) and 1P1S (bottom) configurations. Median values are indicated by the vertical dotted lines. For planetary (stellar) eccentricities, the initial Rayleigh (uniform) distribution is shown as a mauve dotted line and a uniform (thermal) distribution as a grey dashed line for reference. Note that in the 2P case the initial eccentricity distributions are not visible due to overlap with the field distributions.
\label{fig:cj_comp_eccentricity}}
\end{figure*}

Beyond the increased rate of more extreme dynamical outcomes, the influence of the collisional stellar environment is also reflected in the properties of surviving systems. As shown in the top row of Figure \ref{fig:sma_ecc}, a small fraction of single-planet systems is excited to high eccentricities as a result of stellar flybys, with less than 3~per cent of all 1P systems attaining eccentricities exceeding 0.2. Similarly, since our initial conditions for the 2P case satisfy the \citet{petrovich_stability_2015} stability criterion, strong planet-planet scattering is initially precluded and any subsequent dynamical excitation therefore results from external perturbations propagating through the system. As shown in the middle row of Figure \ref{fig:sma_ecc} and the top row of Figure \ref{fig:cj_comp_eccentricity}, while the outer planet is frequently excited to high eccentricities by stellar flybys, this excitation is not efficiently propagated to the inner orbit. Approximately 18~per cent of surviving outer planetary companions reach eccentricities in excess of 0.2, compared to just 3~per cent of all inner planets. As a result, the median final eccentricity of surviving inner CJs in the cluster sample remains low at $\sim$0.06, statistically indistinguishable from the control sample and the initial Rayleigh distribution. In contrast, in the 1P1S configuration, the median eccentricity of surviving CJs in the control sample increases to 0.15, driven by initially eccentric and mutually inclined companions, whereas for the overall cluster sample this is further elevated to 0.33, as shown in the bottom left panel of Figure \ref{fig:cj_comp_eccentricity}.

Among the surviving stellar companions, the influence of the cluster environment is similarly evident in the deviation from the initial uniform eccentricity distribution (bottom right panel of Figure \ref{fig:cj_comp_eccentricity}). It should be noted that the very highest eccentricities are already initially excluded by our choice of stability criteria, leading to a slight depletion in the field sample relative to the uniform distribution. However, the companion eccentricities in the cluster are clearly pumped to preferentially higher values by close encounters, reaching a median value of 0.62, compared to 0.46 in the field. Furthermore, the eccentricity distribution of surviving stellar companions in the cluster approaches a thermal distribution ($f(e) = 2e$), consistent with the expectation that repeated gravitational scattering will drive any bound population toward a thermally relaxed state \citep{heggie_gravitational_2003}.

A further consequence of the dynamical excitation of cold Jupiters in the cluster environment concerns the survival of any inner terrestrial planets that may co-exist in these systems. Based on numerical simulations, \citet{becker_warm_2026} found that an inner terrestrial planet can survive tidal migration of an outer giant only if the mutual orbital spacing in units of the mutual Hill radius
\begin{equation}
    \Delta_{\rm Hill}^{\rm peri} = \left( \frac{\mu_{\rm in} + \mu_{\rm out}}{3} \right)^{1/3} \frac{a_{\rm in} + a_{\rm out}(1 - e_{\rm out})}{2},
\end{equation}
exceeds $\sim$14. We take $m_{\rm in} = 1~{\rm M_\oplus}$ and $a_{\rm in} = 0.02$~AU, corresponding to the innermost known companion to a hot Jupiter host (WASP-84; \citealt{maciejewski_hot_2023}), and use the minimum periapsis distances of surviving CJs recorded during our simulations to estimate the fraction of disrupted inner terrestrial planets. We find that $\sim$9~per cent of surviving CJs in the 1P1S cluster sample would destabilize a co-existing terrestrial planet, compared to $\sim$4~per cent in the field. These disruption fractions increase to $\sim$32 and $\sim$25~per cent, respectively, if the inner terrestrial planet is placed at 0.1~AU instead. Regardless of the terrestrial planet's orbital radius, such planets would be disrupted in all HJ-forming systems during the high-eccentricity migration phase in both cluster and field environments, consistent with the paucity of inner companions to hot Jupiters \citep{mustill_destruction_2015, sha_occurrence_2026}.

\section{Discussion}
\label{sec:discussion}

\subsection{Occurrence Rate Estimates}
\label{sec:occurrencerate}

\begin{figure}
\includegraphics[width=\columnwidth]{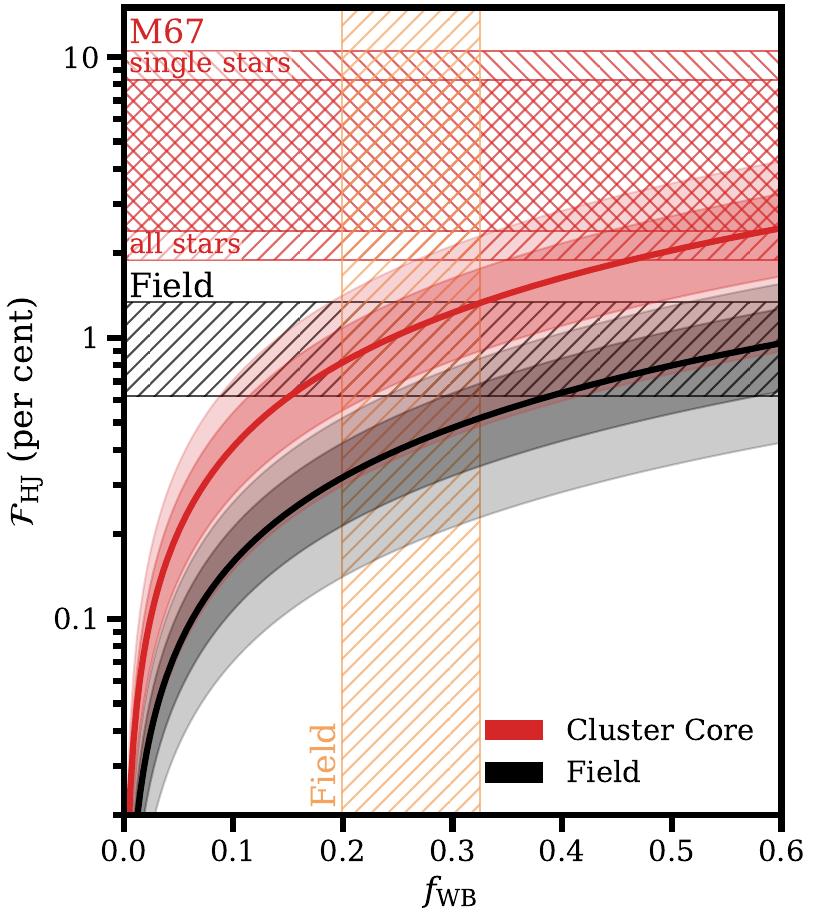}
\caption{Predicted hot Jupiter occurrence rate $\mathcal{F}_{\rm HJ}$ as a function of the wide binary fraction $f_{\rm WB}$ for planetary systems in the cluster core ($r < r_{\rm h}$, red) and the Galactic field (black) based on the 1P1S simulation sample. The solid lines indicate the fiducial estimate with $f_{\rm surv} = 0.6$ based on \citet{weldon_saving_2026}, and the dotted lines show the predicted $\mathcal{F}_{\rm HJ}$ in the absence of partial survival of tidally disrupted systems ($f_{\rm surv} = 0$). The inner shaded bands correspond to the propagated uncertainty on the fiducial $\mathcal{F}_{\rm HJ}$ from $f_{\rm GP}$ \citep{hirsch_understanding_2021} and our model output ($f_{\rm HJ}$, $f_{\rm TD}$), while the outer shaded bands show the combined uncertainty range for $f_{\rm surv} \in [0, 1]$. The hatched bands correspond to the observed HJ occurrence rate in M67 (red; \citealt{thomas_search_2024}) and the Galactic field (black; \citealt{beleznay_exploring_2022}), and the binary fraction of systems hosting giant planets in the field (orange; \citealt{rosenthal_california_2023}), all with their corresponding 1$\sigma$ uncertainties.
\label{fig:hj_occurrence_rate}}
\end{figure}

\begin{figure*}
\includegraphics[width=\textwidth]{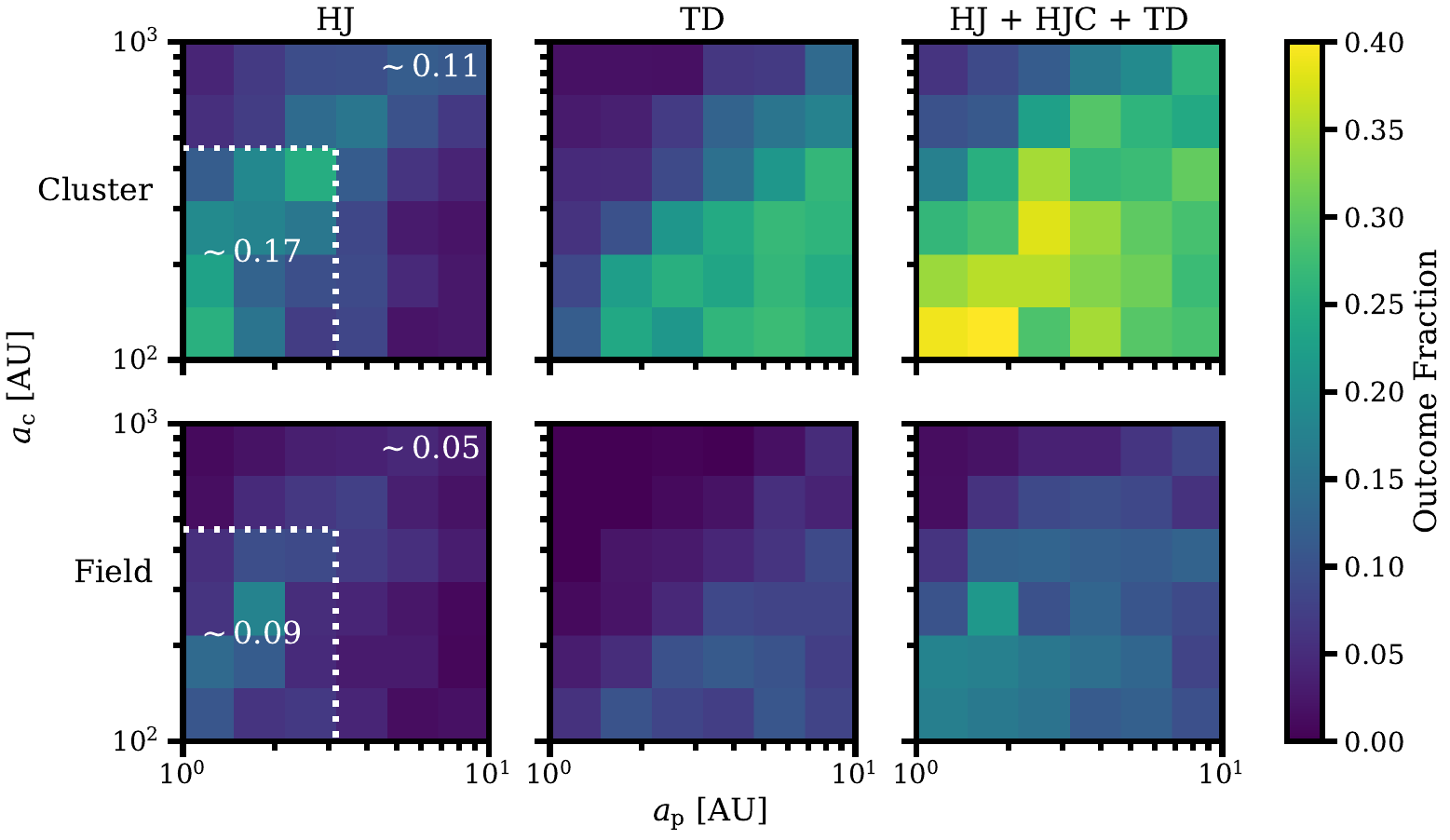}
\caption{Hot Jupiter (HJ, left), tidal disruption (TD, middle), and optimistic hot Jupiter (HJ + HJC + TD, right) yields as a function of the initial inner and outer semi-major axes in the cluster (top) and field (bottom) samples for the 1P1S configuration. The transition from HJ formation to tidal disruption as the dominant outcome is clearly visible around $a_{\rm p} \sim 3$~AU. Integrated over the full parameter space, the cluster environment enhances the HJ yield by a factor of $\sim$2 relative to the field, as indicated by the numbers in the top-right corner of the left-hand panels. A similar enhancement is present in the highlighted subset but the corresponding baseline HJ yield in the field is also approximately doubled compared to the full sample.
\label{fig:hj_fraction}}
\end{figure*}

To connect our results with the observed demographics in M67, we must estimate the primordial fraction of giant planet systems in configurations matching our initial conditions. In general, giant planets with properties similar to the inner planet in our simulations are found around $\sim$10~per cent of all FGK stars \citep{mayor_harps_2011, fernandes_hints_2019, fulton_california_2021}, although surveys of giant planets within 10~AU among the nearest Sun-like stars point to an occurrence rate nearer to $\sim$20~per cent \citep{hirsch_understanding_2021}. Due to detectability limits, the rate of distant planetary companions is not well-constrained, although direct imaging surveys point to an occurrence rate of order 5--10~per cent for substellar companions between 10--300~AU \citep{nielsen_gemini_2019, vigan_sphere_2021}. Upcoming data releases from Gaia are expected to shed more light on the population of wide planetary companions \citep{lammers_exoplanet_2025}. Furthermore, an analysis of the California Legacy Survey by \citet{rosenthal_california_2023} found that at least a third of all stars hosting giant planets host more than one of them. Finally, $\sim$26~per cent of all giant planet systems are in stellar binaries \citep{rosenthal_california_2023}, in line with the overall field rate of systems in wide binaries \citep{raghavan_survey_2010}.

The corresponding occurrence rates for hot Jupiters yield clues as to their dominant formation pathways. While HJs generally lack nearby companions \citep{wu_evidence_2023}, previous work suggests $70 \pm 8$~per cent of HJ systems also host an outer planet with mass 1--13~$\rm M_{J}$ within 1--20~AU \citep{knutson_friends_2014, bryan_statistics_2016}. More recently, \citet{zink_hot_2023} obtained a lower limit of $\sim$30~per cent for the fraction of HJ systems which host an outer giant planet. Outer stellar companions are also relatively common: \citet{ngo_friends_2016} found $47 \pm 7$~per cent of HJ-hosting systems to also host a stellar companion at 50--2,000~AU, although only a fraction ($16 \pm 5$~per cent) of these binary configurations are compatible with efficient ZLK oscillations. Notably, \citet{stephan_two_2024} suggest that even seemingly solitary hot Jupiters could have formed in binary systems that were subsequently disrupted by white dwarf kicks, implying the true primordial binary fraction among HJ progenitors may be higher than present-day observations suggest. More generally, these observed frequencies only pertain to surviving planetary and stellar companions, and the primordial companion fraction -- especially in a dense cluster environment -- could have been substantially higher.

Based on these occurrence rates, some simple estimates can be made. Focusing on the stellar companion (1P1S) case, if we adopt a wide ($a_{\rm bin} \gtrsim 100$~AU) binary fraction of $f_{\rm WB} = 30$~per cent and a giant planet occurrence rate of $f_{\rm GP} = 20^{+7}_{-6}$~per cent in such binaries from \citet{hirsch_understanding_2021}, the expected HJ rate from the binary-mediated channel would be
\begin{equation}
\begin{split}
\label{eq:hj_rate_nosurvival}
    \mathcal{F}_{\rm HJ} &= f_{\rm GP} \times f_{\rm WB} \times f_{\rm HJ}\\ &= 0.20 \times 0.30 \times 0.11 \\ &= 0.7 \pm 0.2~{\rm per~cent}
\end{split}
\end{equation}
in the core of the cluster.\footnote{In evaluating this and all subsequent occurrence rate estimates, we propagate symmetrized uncertainties on $f_{\rm GP}$, $f_{\rm HJ}$, and $f_{\rm TD}$ while treating $f_{\rm WB}$ and $f_{\rm surv}$ as free parameters (Figure \ref{fig:hj_occurrence_rate}). The intermediate values of our outcome yields are rounded to two significant figures for clarity.} Thus, factoring in realistic occurrence rates of giant planets and outer stellar companions, we find that the flyby-induced HEM pathway falls short of the high HJ incidence in M67. However, the resulting HJ yield in the control sample ($0.3 \pm0.1$~per cent) is similarly too low to match the observed occurrence rate of $\sim$1~per cent in the Galactic field \citep{beleznay_exploring_2022}.

It has previously been recognized that relaxing the tidal disruption criterion (Equation \ref{eq:td_limit}) with $\eta = 2.7$ established numerically by \citet{guillochon_consequences_2011} can boost the hot Jupiter yield by allowing systems that undergo close periapsis passages to survive as HJs \citep{naoz_formation_2012, petrovich_steady-state_2015, teyssandier_formation_2019}. Recent work on HJ formation in stellar binaries by \citet{weldon_saving_2026} has demonstrated that incorporating a realistic prescription for partial mass loss results in the survival of a substantial fraction of systems that would otherwise be considered tidally disrupted based on the fixed \citet{guillochon_consequences_2011} criterion. Based on their numerical experiments, \citet{weldon_saving_2026} conclude that ZLK induced by a stellar binary companion may be prevalent enough to explain the HJ occurrence rate around field FGK stars if partial mass loss during close periapsis passages is adequately accounted for. Recent hydrodynamical simulations by \citet{fan_where_2026} similarly demonstrate the role of the planetary internal structure in determining the outcome of high-eccentricity migration, with the presence of a rocky core decreasing the radius at which a gas giant is tidally disrupted.

Although we do not include a treatment of mass loss in our model, we can incorporate its effect on the resulting HJ yield by refining our estimate of the HJ occurrence rate in the cluster core above as
\begin{equation}
\begin{split}
\label{eq:hj_rate_cons}
    \mathcal{F}_{\rm HJ} &= f_{\rm GP} \times f_{\rm WB} \times (f_{\rm HJ} + f_{\rm TD} \times f_{\rm surv})\\ &= 0.20 \times 0.30 \times (0.11 + 0.16 \times 0.60) \\ &= 1.2 \pm 0.4~{\rm per~cent}
\end{split}
\end{equation}
where the fiducial value for the fraction of tidally disrupted systems that survive as hot Jupiters ($f_{\rm surv} = 0.6$) is based on the results of \citet{weldon_saving_2026}. Accounting for the survival of these partially disrupted systems nearly doubles the effective hot Jupiter yield from this pathway. Without this enhancement, the predicted occurrence rates fall short of the observed fractions in both environments for any realistic wide binary fraction. We also caution that this value of $f_{\rm surv}$ is not directly applicable to our sample, since \citet{weldon_saving_2026} adopt different initial conditions -- including sampling both stellar and planetary masses and drawing companion semi-major axes from a log normal distribution as suggested by surveys in the field \citep{duquennoy_multiplicity_1991, raghavan_survey_2010} -- which affect the distribution of periapsis distances and therefore the survival fraction. We illustrate how the uncertainty in the survival probability $f_{\rm surv}$ affects our estimate of the predicted HJ occurrence rate in Figure \ref{fig:hj_occurrence_rate}. A fully self-consistent estimate would require implementing a partial mass-loss prescription in our model, which we defer to future work.

While our choice of $f_{\rm WB} = 0.3$ is motivated by the observed binarity of giant planet hosts in the field as well as the present-day binary fraction in M67, this likely underestimates the primordial value given the low survival rate of wide binaries in a dense cluster environment. Indeed, both the $N$-body model of M67 by \citet{hurley_complete_2005} and the Monte Carlo counterpart by \citet{giersz_monte_2008} adopt a primordial binary fraction of 0.5 to reproduce the observed properties of the cluster. While the orbital separation of primordial binaries in both models was only sampled up to 50~AU to exclude soft binaries, the strong correlation between the close and wide binary fractions across a range of stellar masses (Table 1 of \citealt{offner_origin_2023}) justifies extrapolating this value to wider systems. Similarly high binary fractions have also been assumed in previous demographic estimates \citep{vick_chaotic_2019}. Hence, adopting the more optimistic yet observationally motivated value of $f_{\rm WB} = 0.5$ yields
\begin{equation}
\begin{split}
\label{eq:hj_rate_opt}
    \mathcal{F}_{\rm HJ} &= 0.20 \times 0.50 \times (0.11 + 0.16 \times 0.60) \\
    &= 2.1 \pm 0.7~{\rm per~cent}
\end{split}
\end{equation}
in the core of the cluster and $0.8 \pm 0.3$~per cent in the field -- in good agreement with the observed Galactic field HJ rate \citep{beleznay_exploring_2022} and consistent with the M67 observations within observational uncertainties \citep{thomas_search_2024}. The dependence of the predicted HJ occurrence rate on both $f_{\rm WB}$ and $f_{\rm surv}$ is shown in Figure \ref{fig:hj_occurrence_rate}, illustrating the need for both a high primordial wide binary fraction and a high survival probability in order to simultaneously reproduce the cluster and field occurrence rates from a single set of initial conditions.

We caution that this final value should be interpreted as an illustrative rather than a self-consistent estimate, since $f_{\rm WB}$ is correlated with $f_{\rm CB}$ \citep{offner_origin_2023} which affects both the encounter rate and perturber sampling in our model and therefore the outcome fractions $f_{\rm HJ}$ and $f_{\rm TD}$ themselves (Section \ref{sec:flybys}). Given the larger cross-sections of binary perturbers, a higher overall binary fraction would increase the perturbation rate experienced by planetary systems \citep{shara_dynamical_2016}, likely boosting both outcome fractions and causing our estimate of $\mathcal{F}_{\rm HJ}$ to be on the conservative side. However, as discussed in Appendix \ref{sec:binaryfraction}, increasing the close binary fraction to 0.5 has little impact on the outcome statistics when truncating the binary population to $a_{\rm bin} < 10$~AU. Strictly speaking, a high primordial fraction of wide binaries would also lead to occasional encounters with other such systems, an effect not accounted for in our model. However, beyond a sufficiently large orbital radius, such encounters are expected to be largely indistinguishable from those with single perturbers, so we do not expect this effect to be dominant.

It should also be emphasized that our choice of initial conditions does not strictly reflect the observed properties of planetary systems. In particular, our assumption of an initially log uniform distribution in $a_{\rm bin}$ does not match observations in the field \citep{duquennoy_multiplicity_1991}. This is a deliberate choice, given our goal of systematically mapping the effect of the cluster environment across the parameter space. The choice of a log normal distribution peaking around $\sim$50~AU in line with observations of solar-type binaries \citep{raghavan_survey_2010} would preferentially produce closer binaries capable of driving more efficient eccentricity excitation, therefore leading to a further increase in the HJ and TD rates. However, the significant uncertainties in the mapping between present-day and primordial stellar and planetary populations preclude the use of observed demographics, which have already been sculpted by dynamical processing, as a direct basis for initial conditions. The primordial binary populations in dense clusters and the field may also differ substantially, and the picture is further complicated by the production of dynamically formed binaries through tidal capture and three-body interactions \citep{heggie_gravitational_2003}, which supplement the primordial population.

These considerations present another possibility for interpreting our results. Rather than trying to directly reproduce the high HJ occurrence rate in M67 using a set of observationally motivated initial conditions, we can benchmark our results by assuming that HEM in binary systems accounts mostly or wholly for the field HJ rate, as suggested by both observations \citep{rice_origins_2022} and theoretical modelling \citep{weldon_saving_2026}. By selecting a subset of the parameter space of initial conditions for which this is true and further assuming that the primordial planet population in the cluster is similar to that in the field, we may evaluate the factor by which HJ formation is enhanced by the collisional environment. As discussed in the context of Table \ref{table:outcomes}, this environmental enhancement factor is $\sim$2 for HJs and $\sim$3 for TDs when integrated over the entire range of initial conditions. However, as shown in Figure \ref{fig:hj_fraction}, this enhancement is not uniform across the parameter space of initial semi-major axes. In particular, while the efficiency of HJ formation is boosted most significantly among systems with the inner planet initially within $\sim$3~AU and the outer stellar companion within $\sim$500~AU, the range of initial conditions for which similar HJ fractions is achieved is also extended. On the other hand, planets initially beyond $\sim$3~AU are predominantly tidally disrupted regardless of the environment.

As a further illustrative example, we consider systems with the inner planet initialized between 1--3~AU and the outer stellar companion between 100--500~AU. This range is broadly consistent with the observational evidence that planet formation is only suppressed in binaries closer than $\sim$100~AU \citep{hirsch_understanding_2021, sullivan_first_2025, venturini_pairs_2026}, whereas wider companions are too distant to efficiently drive ZLK oscillations. In our control sample, systems in this parameter range form HJs at a rate $\sim$2 times higher than the sample-averaged rate (Figure \ref{fig:hj_fraction}). Assuming this subset broadly accounts for the dominant binary-driven HEM contribution to the field population, the corresponding cluster enhancement of a factor of $\sim$2 in the same region of parameter space brings the inferred M67 HJ occurrence rate broadly in line with the observations, although the precise value is sensitive to the adopted binary fraction and tidal model, as discussed above. This exercise demonstrates that rather than requiring unusual initial conditions or an anomalously high primordial planet fraction in M67, the cluster environment alone can plausibly account for the observed excess if binary-driven HEM is assumed to dominate in both the cluster and the field.

However, we note that alternative formation channels (Section \ref{sec:alternativechannels}) are also likely to contribute a fraction of the HJ population in both environments. In particular, planet-planet interactions may be responsible for the formation of some hot Jupiters, as suggested by the presence of outer planetary companions to known HJs \citep{knutson_friends_2014, bryan_statistics_2016, zink_hot_2023}. Adopting the giant planet occurrence rate of $f_{\rm GP} = 18^{+4}_{-3}$~per cent around single stars from \citet{hirsch_understanding_2021}\footnote{Strictly speaking, the occurrence rate estimates by \citet{hirsch_understanding_2021} are for giant planets between 0.1--10~AU rather than our sampling range of 1--10~AU. However, given that many of these close-in planets have likely migrated from beyond $\sim$1~AU, we argue that this value better reflects the primordial fraction of systems hosting cold giants.} and assuming that a third of these also host an outer giant \citep{rosenthal_california_2023}, the contribution from two-planet (2P) systems to the HJ occurrence in the cluster core comes out to be $\mathcal{F}_{\rm HJ} \sim 0.06$~per cent, where we assume a similar survival rate of tidally disrupted systems as in the 1P1S case. This is substantially lower than the contribution from the stellar companion channel, reflecting both the lower sensitivity of two-planet systems to the cluster environment and the conservative stability criteria imposed in sampling our initial conditions (discussed in Section \ref{sec:clusterfield}). We leave a detailed analysis of the combined contribution of all environmentally enhanced channels for future work. Nevertheless, the negligible contribution from 2P systems relative to the stellar companion channel and the consistency of a single set of initial conditions with both the observed field and cluster occurrence rates point to binary-driven HEM as the dominant formation pathway across environments, broadly consistent with the observational evidence from obliquity measurements \citep{rice_origins_2022}.

Finally, we note that the HJs formed in our simulations have very short final orbital periods ($P \lesssim 5$~days), a characteristic feature of tidal migration driven by stellar binary ZLK \citep{petrovich_steady-state_2015, anderson_formation_2016}. This presents a potential tension with the observed orbital periods (4--7~days) of the HJs discovered in M67 \citep{brucalassi_search_2016}. However, these planets also have minimum masses significantly lower ($m_{\rm p} \sin{i} \approx 0.32-0.46~{\rm M_{J}}$) than the fixed 1~$\rm M_J$ we chose for the inner planet in our simulations. The tendency of lower-mass and therefore lower-density giant planets to circularize at longer-period orbits \citep{anderson_formation_2016} may therefore partially reconcile this discrepancy. Additionally, the adoption of more sophisticated tidal prescriptions may allow circularization to stall at wider orbits, broadening the final period distribution. In particular, the inclusion of chaotic tides has been shown to produce a significantly longer tail in the hot Jupiter period distribution, an effect that is particularly pronounced for lower-mass giant planets \citep{vick_chaotic_2019}. Quantifying the effect of alternative tidal prescriptions or a distribution of inner planet masses and radii (as opposed to the fixed 1~$\rm M_{J}$ and 1~$\rm R_{J}$ used here) is beyond the scope of this work.

\subsection{Effect of Cluster Evolution}
\label{sec:cluster_evolution}

The evolutionary history of clusters such as M67 allows us to probe how various cluster-scale effects shape planetary demographics among both bound and escaping populations. Based on their $N$-body model, \citet{hurley_complete_2005} hypothesize that M67 has lost $\sim$90~per cent of its mass over its lifetime. Using data from Gaia EDR3, \citet{alvarez-baena_longevity_2024} similarly found that the cluster has lost $\sim$60~per cent of its birth mass and has a cuspy core indicative of an advanced stage of dynamical evolution. The Galactic orbit and internal kinematics of M67 further suggest that the cluster is undergoing periodic ($\sim$40~Myr) crossings of the Galactic disc which inhibit dynamical relaxation \citep{carrera_extended_2019, zwitter_galah_2021}. We note that this effect was not included in previous cluster models by \citet{hurley_complete_2005} and \citet{giersz_monte_2008} used as a basis for our work.

M67 has also been found to exhibit significant mass segregation \citep{gao_machine-learning-based_2018, carrera_extended_2019}. As a result, the binary population in the cluster is more centrally concentrated compared to single main sequence stars \citep{geller_stellar_2015}, with the fraction of binaries out to periods of $10^4$ days reaching $\sim$70~per cent in the core \citep{geller_stellar_2021}. As two-body relaxation and tidal stripping preferentially remove low-mass stars and systems residing at large cluster-centric radii, the core becomes increasingly dominated by massive and dynamically hardened systems \citep{spitzer_equipartition_1969}.\footnote{While our fiducial model draws perturber masses from a global initial mass function (Equation \ref{eq:imf}), we performed additional experiments (detailed in Appendix \ref{sec:perturbermass}) using a fixed perturber mass to constrain the potential impact of mass segregation on our outcome statistics.} Planetary systems retained in the cluster over Gyr timescales are therefore also those that have undergone the most dynamical processing. Cluster dissolution thus introduces a form of retention bias, whereby the present-day bound population is enriched in dynamically processed systems, while escapers retain a comparatively larger fraction of unperturbed cold Jupiters.

While most systems in our simulation remain bound to the cluster over the 4~Gyr integration, as expected given that the 1~$\rm M_\odot$ host stars are more massive than the average cluster member and therefore less susceptible to tidal stripping, the dynamically processed population of escaping stars may nevertheless have implications for planetary demographics in the Galactic field. In particular, the HJ and TD outcome fractions among cluster escapers are enhanced by of order 33 and 45~per cent, respectively, for the 1P1S configuration, compared to isolated systems in the field (Table \ref{table:outcomes}). In terms of the overall HJ occurrence rate evaluated using Equation \ref{eq:hj_rate_opt}, this translates to $1.1 \pm 0.4$~per cent, representing a modest $\sim$35~per cent enhancement relative to the control sample with no flybys. The rate at which HJ-forming systems escape the cluster is roughly constant with time, with half of the escaping population having done so by $\sim$2.7~Gyr. Since most stars in the Galaxy are thought to have formed in clustered environments \citep{lada_embedded_2003}, this suggests that a non-negligible fraction of the field HJ population may owe their existence to dynamical processing in their birth clusters.

In our simulations, the majority of HJ-forming and tidally disrupted systems that ultimately escape into the field undergo migration while still bound to the cluster. In these cases, the cluster environment directly drives or accelerates the dynamical pathways leading to orbital shrinkage. However, a fraction of systems only complete their migration after escaping into the field. In such cases, cluster-induced perturbations act to prime the orbital architectures, exciting eccentricities or mutual inclinations to values that subsequently enable secular evolution and tidal dissipation to operate in isolation. Indeed, it has been suggested that certain planetary systems with anomalous inclinations could have been perturbed in clusters before escaping \citep{cai_signatures_2018}. The identification of field planetary systems that may originate in cluster environments has previously been attempted based on kinematic signatures \citep{winter_stellar_2020}. However, these efforts are complicated by the need to disentangle Galactic kinematic signatures from intrinsic host star properties \citep{adibekyan_stellar_2021, mustill_hot_2022, blaylock-squibbs_evolution_2023, blaylock-squibbs_no_2024, kontiainen_hot_2025}.

\subsection{Comparison with Previous Work}

We now turn to comparing our results with previous work on the topic. For single-planet (1P) systems, the stellar number density in our model of M67 is too low and close encounters therefore too infrequent to drive efficient eccentricity diffusion. The central stellar number density in our M67 model at the beginning of the simulation is approximately 260~$\rm pc^{-3}$, declining to 35~$\rm pc^{-3}$ by the end of the integration. This is roughly two orders of magnitude lower than the densities of ${\sim} 10^4~{\rm pc^{-3}}$ at which \citet{hamers_hot_2017} and \citet{wirth_hot_2025} found direct scattering to produce HJ formation probabilities of order 2~per cent, accompanied by a similar tidal disruption rate. Direct scattering is therefore a viable HJ formation pathway only in denser globular clusters, although the elevated binary fraction in open clusters may partially compensate for this deficit.

In the two-planet (2P) case, our HJ and TD yields in the cluster are in line with the results of \citet{shara_dynamical_2016} who simulated the evolution of systems with two Jovian planets in a star cluster using direct $N$-body techniques. They found hot Jupiters to be formed at a frequency of $\sim$1~per cent of all planetary systems in a cluster similar to M67. While the authors did not implement a tidal disruption criterion, it is worth pointing out that of the three HJs formed in their simulations, two should be considered tidally disrupted based on the \citet{guillochon_consequences_2011} criterion while the surviving one subsequently escapes the cluster. Our results similarly point to planets being more likely to be tidally disrupted than being circularized as hot Jupiters by a factor of a few, although the ratio is sensitively dependent on the precise treatment of tidal evolution.

Our findings are also consistent with those of \citet{wang_hot_2020} who estimated a lower limit to the cluster-driven HJ formation rate by considering only close encounters capable of changing the semi-major axis ratio of two-planet systems. In their model of M67, they found an occurrence rate approximately 50 times smaller than observed, leading them to conclude that direct scattering in two-planet systems is insufficient to account for the high incidence in the cluster. This result does not depend sensitively on the encounter rate $\Gamma$, as the formation bottleneck is not the frequency of sufficiently disruptive flybys but rather the narrow parameter space in which a planet survives the initial scattering event and subsequently migrates to a stable HJ orbit without being tidally disrupted or ejected.

\citet{rodet_correlation_2021} similarly studied the excitation of initially coplanar two-planet systems into the ZLK window through extensive $N$-body experiments, finding that a flyby must approach to within a distance comparable to the semi-major axis of the outer companion in order to have a significant dynamical impact. Inclination rather than eccentricity excitation therefore presents the main obstacle to HJ formation via ZLK in two-planet systems, in line with earlier results from \citet{li_cross-sections_2015} and \citet{wang_hot_2020}, and qualitatively consistent with our results. Based on their simulations, \citet{rodet_correlation_2021} further develop a semi-analytic estimate of the resulting HJ yield (see their Equation 20). Taking the average perturber mass and companion orbital separation from our model, while adopting their fiducial values for the other parameters, we integrate their expression over $n_\star$ and $\sigma_v$ along the cluster half-mass radius $r_{\rm h}$ to estimate the total expected HJ yield from this channel. The resulting estimate of $\sim$0.4~per cent is comparable to our results for the 2P case when the potential survival of a fraction of seemingly tidally disrupted systems is taken into account.

Finally, in the stellar companion (1P1S) case, \citet{li_making_2024} studied the role of the collective cluster potential in modifying the companion star's orbit to be more favourable to HJ formation through extreme ZLK (XZLK). They found the fraction of XZLK-activated systems (classified as those with periapsis distance below 0.022~AU) in the cluster environment to be boosted by a factor of $\sim$4 compared to the field. This is in line with the difference between the TD rates in the cluster and field samples in our 1P1S configuration. It is worth noting that \citet{li_making_2024} did not include a tidal prescription in their model, instead assuming that $1/3$ of all XZLK-activated systems form HJs. The resulting HJ occurrence rate was still found to fall short of that observed in M67, suggesting multiple HJ formation channels must act in concert to produce the observed demographics.

\subsection{Alternative Formation Pathways}
\label{sec:alternativechannels}

Based on studies of the obliquity distribution of hot Jupiter host stars, HEM appears to be the dominant formation pathway at the population level \citep{rice_origins_2022}. As discussed in Section \ref{sec:occurrencerate}, our predicted occurrence rates support this picture. However, while environmental effects in dense clusters open new migration pathways via dynamical excitation, they may also limit the primordial reservoir of CJs. In particular, external photoevaporation \citep{winter_growth_2022}, ambient background heating \citep{ndugu_planet_2018}, and disc truncation due to early-time stellar flybys \citep{ndugu_planet_2022} may potentially suppress the formation of cold Jupiters that would later serve as HJ precursors, although the latter effect is likely only significant in environments much denser than M67 \citep{winter_protoplanetary_2018}. Specifically, \citet{pfalzner_did_2018} find that stellar flybys can effectively truncate more than 26~per cent of discs to sizes $\lesssim$30~AU in the open cluster M44. However, based on the cluster parameters used as the basis for their $N$-body model, the mean stellar number density within the half-mass radius reaches ${\gtrsim}10^5~{\rm pc^{-3}}$, three orders of magnitude higher than the corresponding value (${\sim}100~{\rm pc^{-3}}$) in the \citet{hurley_complete_2005} model of M67. Truncation by stellar flybys is hence unlikely to significantly modulate the disc and exoplanet populations in the cluster.

In realistic star-forming environments, disc truncation by external photoevaporation generally dominates over tidal encounters in the embedded phase \citet{winter_protoplanetary_2018}. Based on our adopted values and the relationship $G_0 = 10^3 (n_\star /{\rm pc^{-3}})^{1/2}$ between the far-ultraviolet (FUV) flux and the local number density from \citet{winter_protoplanetary_2018}, we find the FUV flux in the core of M67 to have reached $G_0 \sim 10^4$ in the early evolutionary stages. However, based on 1D disc evolution models including external photoevaporation, \citet{sellek_evolution_2020} found the planet formation efficiency at the snowline not to be severely affected even in the presence of such a strong FUV background provided the formation process is relatively rapid, as is expected to be the case for giant planets \citep{ikoma_formation_2025}. Shielding by the giant molecular cloud may further suppress the effect of external FUV fields \citep{qiao_evolution_2022} in these early stages. Indeed, the presence of planets in M67 suggests giant planet formation is not strongly suppressed by the radiative environment. On the other hand, quantifying the extent to which interactions during the embedded phase may contribute to the in-situ formation of close-in giant planets (see \citealt{thies_natural_2011}) is beyond the scope of this work.

Disc migration likely contributes only a limited fraction of the HJ population \citep{kawai_identifying_2025, schmidt_most_2026}. Based on a sample of hot Jupiters observed by TESS, \citet{sha_occurrence_2026} place a lower limit of $7.6^{+5.5}_{-3.8}$~per cent on the fraction of HJs with nearby planetary companions which could not have survived HEM (Section \ref{sec:dynamicalsculpting}; see also \citealt{mustill_destruction_2015, becker_warm_2026}). Further accounting for the long-term survival of such companions, this translates to $\sim$20~per cent of HJs in the field having formed via disc migration. Given that disc migration should operate equally efficiently in both dense cluster environments and the Galactic field, thereby providing a constant baseline contribution to $\mathcal{F}_{\rm HJ}$, one would expect the relative fraction of HJs formed via HEM to be higher in clusters. This could in principle be inferred from stellar obliquity measurements \citep{rice_origins_2022, albrecht_stellar_2022}, since disc-migrated planets are expected to be broadly aligned with their host stars. However, the limited size of the sample and the non-transiting nature of the confirmed HJs in M67 do not currently permit such a comparison.

Finally, in this work we have chosen to focus on systems with only one additional companion. However, several higher-multiplicity pathways may contribute to the HJ population, potentially with enhanced efficiency in cluster environments. In particular, in systems of higher planetary multiplicity, secular chaos triggered by flybys may lead to more efficient HJ formation \citep{wang_hot_2022}, although both population synthesis studies such as the Bern model \citep{emsenhuber_new_2021} and observational surveys \citep{li_intrinsic_2026} suggest that giant planets form at most in pairs rather than in higher multiples.\footnote{Notable exceptions to this include the well-studied HR 8799 system \citep{marois_direct_2008, marois_images_2010} as well as the recently discovered TOI-375 system consisting of a hot Jupiter and two outer giant planets \citep{reinarz_transiting_2026}.} We have also neglected the possibility of "double" HJ formation in binary systems in which both components initially host a cold Jupiter \citep{liu_double_2026}. Finally, the presence of a tertiary stellar companion may also contribute to HJ formation via the recently proposed "eccentricity cascade" mechanism \citep{yang_third_2025, yang_analytical_2025} which could be activated by stellar flybys.

\subsection{Caveats and Future Work}
\label{sec:caveats}

We emphasize that our model is highly simplified so as to allow the efficient treatment of the multiple physical scales involved. We have also adopted an agnostic set of initial conditions in order to probe the effect of the cluster environment across much of the parameter space. A natural extension of this work would be to apply importance sampling to re-weight the outcome fractions post hoc using more physically motivated initial condition distributions, such as those from planet formation population synthesis models \citep{emsenhuber_new_2021}. This would allow the parameter space explored here to be mapped onto observationally motivated priors without the computational expense of additional simulation runs, provided the existing sample provides adequate coverage of the relevant parameter space. For these purposes, the sampling methodology could be extended to cover a range of HJ progenitor and host masses and radii.

Among the main theoretical uncertainties in modelling hot Jupiter formation is the prescription for the tidal evolution of migrating HJ progenitors. In the future, more realistic tidal models could be incorporated into our model. A self-consistent implementation of chaotic tides was recently introduced in the \texttt{REBOUNDx} framework by \citet{liveoak_self-consistent_2025}. However, in this work, we have chosen to use the hybrid tidal model from \citet{moe_dynamical_2018} and \citet{rozner_inflated_2022} to ensure consistency between the tidal implementations during periods of secular and non-secular evolution. We have also neglected the possible role played by radius inflation during tidal migration \citep{rozner_inflated_2022, glanz_inflated_2022}.

Furthermore, while the global parameters of the cluster evolve over time, we assume it retains a Plummer-like density distribution throughout. This likely underestimates the stellar density in the core, particularly at late times when mass segregation and two-body relaxation produce a cuspier central structure \citep{alvarez-baena_longevity_2024}. A more detailed treatment would therefore increase the encounter rate for core members relative to our model, suggesting our HJ and TD yields for the $r < r_{\rm h}$ subsample may be on the conservative side. Other factors currently neglected include the evolution of the perturber population, particularly mass segregation \citep{gao_machine-learning-based_2018} and the production of dynamically formed binaries, primordial cluster substructure \citep{parker_effects_2012, parker_dependence_2023}, and the effect of the cluster tidal field \citep{hamilton_secular_2019, hamilton_secular_2019-1}.

Going forward, we will experiment with different cluster evolution prescriptions to better understand potential correlations between planet demographics and the dynamical history of the host cluster. Given the separation of scales, differences in cluster evolution are expected to affect planetary demographics only through variation in the statistical distributions from which flyby properties are sampled rather than any more immediate effects. Examples of clusters our model could potentially be applied to include 47 Tuc \citep{giersz_monte_2011}, M4 \citep{heggie_monte_2008, heggie_towards_2014}, and NGC 6397 \citep{giersz_monte_2009}, of which dedicated evolutionary models already exist, as well as the open clusters to be targeted by the Plato mission \citep{rauer_plato_2025} in the LOPS2 field \citep{nascimbeni_plato_2025}.

\section{Conclusions}
\label{sec:conclusions}

In this work, we develop an efficient hybrid model for simulating the evolution of giant planet systems in an evolving star cluster environment and apply it to study flyby-induced hot Jupiter formation in the open cluster M67. Our main conclusions are as follows:
\begin{itemize}[leftmargin=*]
    \item Flyby-induced HJ formation in single- and two-planet systems is inefficient in open cluster environments with densities comparable to M67. The dominant environmentally driven pathway is instead the perturbation of initially quiescent stellar binaries into configurations that activate ZLK oscillations, driving the inner planet to high eccentricities amenable to tidal capture and circularization.
    \item For planets in stellar binaries, interactions in the cluster environment boost the HJ and TD yields by factors of $\sim$2 and $\sim$3, respectively, relative to isolated evolution in the field. Using observationally motivated estimates of the primordial giant planet and binary populations, and assuming that binary-driven HEM accounts for the field HJ rate, the resulting cluster occurrence rate is consistent with the elevated rate observed in M67 within uncertainties.
    \item The HJ yield depends on the primordial binary fraction, the giant planet occurrence in binaries, and the adopted tidal model. Adopting a primordial wide binary fraction of $\sim$50~per cent -- an extrapolation from $N$-body models of M67 restricted to closer binaries -- and a giant planet occurrence rate of $\sim$20~per cent in such systems reproduces both the field and M67 occurrence rates within uncertainties without further invoking unusual initial conditions.
    \item Only $\sim$1/3 of HJ-forming systems remaining in the cluster retain their outer stellar companion. This reconciles the binary-driven formation pathway with the lack of stellar companions to the hot Jupiter hosts discovered in M67, and further implies that the present-day companion fraction among HJ hosts underestimates the primordial binary fraction of HJ progenitors.
    \item HJ occurrence among systems that escape the cluster is enhanced by $\sim$35~per cent compared to isolated systems, suggesting that dynamical processing in dissolved clusters may contribute a non-negligible fraction of the field HJ population.
\end{itemize}

Our results highlight the role of cluster-driven dynamical processing in shaping the hot Jupiter population in M67-like environments. More broadly, our work underscores the need to interpret exoplanet demographics within the context of their birth environments, as a non-negligible fraction of field systems may carry imprints of early evolution in now-dissolved clusters. Future observational constraints on binary fractions and planet occurrence in open clusters will be critical for testing this framework and for disentangling the role of environmental processing in giant planet evolution.

\section*{Acknowledgements}
We would like to thank the anonymous referee for their helpful comments which improved the quality of this paper. MVK thanks the UK Science and Technology Facilities Council (STFC) for a Ph.D. studentship and also thanks Cristobal Petrovich, Douglas Lin, Roman Rafikov, and Dolev Bashi for useful discussions. AJW has been supported by the Royal Society through a University Research Fellowship grant number URF\textbackslash R1\textbackslash 241791.

\section*{Data Availability}
The simulation code underlying this article will be shared on reasonable request to the corresponding author. The cluster evolution model of M67 used in this work is based on the publicly available $N$-body snapshots of \citet{hurley_complete_2005}, accessible at \href{https://astronomy.swin.edu.au/~jhurley/nbody/archive.html}{https://astronomy.swin.edu.au/~jhurley/nbody/archive.html}.

\bibliographystyle{mnras}
\bibliography{references}

\newpage
\appendix

\section{Impact of Perturber Population}
\label{sec:perturberpopulation}

As described in Section \ref{sec:caveats}, our simplified cluster model neglects a number of physical effects tied to the long-term evolution of the cluster, most prominently the evolution of the binary population and mass segregation. Here, we summarize our findings from additional experiments designed to investigate how the properties of the perturber population affect the outcome statistics.

\subsection{Binary Fraction}
\label{sec:binaryfraction}

To investigate the role of the close binary fraction in determining the extent of dynamical processing, we performed an additional experiment with 5,000 systems in the 1P1S configuration evolving in a cluster with a constant $f_{\rm CB} = 0.5$ across the stellar mass range (Figure \ref{fig:binary_fraction}). This increases the mean perturber mass to $\langle m_{\rm sys} \rangle \approx 0.70~{\rm M_\odot}$, which is correspondingly reflected in a slightly lower stellar number density $n_\star$ to conserve the local mass density $\rho_\star$ in the cluster.

The results of this experiment are largely indistinguishable from our fiducial simulation, with the predicted optimistic HJ occurrence rate (Equation \ref{eq:hj_rate_opt}) in the core remaining nearly unchanged at $2.0 \pm 0.7$~per cent. The only notable difference relative to the fiducial simulation is a slight increase in the fraction of systems escaping the cluster to $\sim$21~per cent, likely reflecting the increased stochastic heating due to the higher average perturber mass. This suggests the exact binarity of the lowest mass perturbers has little impact on the outcome statistics among the bound cluster population. This is to be expected based on our choice of $q$ between 0.1--1 which limits the extent to which binarity can contribute to the total perturbation strength, especially among low-mass primaries. On the other hand, while a constant $f_{\rm CB} = 0.5$ effectively decreases the binary fraction for the most massive perturbers ($m_\star \gtrsim 7~{\rm M_\odot}$) relative to our fiducial model, their rapid evolution off the main sequence (Equation \ref{eq:mainsequence}) limits their interaction window to the first few Myr, rendering this discrepancy negligible over the $\sim$4~Gyr simulation.

\subsection{Perturber Mass}
\label{sec:perturbermass}

To investigate the role of mass segregation and the assumption of a globally sampled initial mass function (Equation \ref{eq:imf}), we performed two additional experiments with 5,000 systems in the 1P1S configuration using fixed-mass perturbers: one with $m_{\rm pert} = 0.66~{\rm M_\odot}$ (matching the mean perturber mass in the fiducial model) and one with $m_{\rm pert} = 1.4~{\rm M_\odot}$ (matching the most massive stars that remain on the main sequence for $\sim$4~Gyr and approximating a dynamically hardened core). To isolate the effect of the perturber mass, we set $f_{\rm CB} = 0$ for both runs. We note that fixing $m_{\rm pert}$ also alters the total encounter rate $\Gamma$. In order to conserve the local cluster mass density $\rho_\star$, increasing the mean perturber mass proportionally lowers the local stellar number density $n_\star$. While the lower number density decreases the geometric encounter rate (Equation \ref{eq:gamma_geo}), the increased total mass of the interacting bodies ($m_{\rm tot}$) simultaneously enhances the gravitationally focused rate (Equation \ref{eq:gamma_focused}).

We find that the heavy, core-like environment slightly suppresses the hot Jupiter yield. In the 1.4~$\rm M_\odot$ scenario, the optimistic HJ occurrence rate in the cluster core (Equation \ref{eq:hj_rate_opt}) drops to $1.7 \pm 0.7$~per cent compared to $1.8 \pm 0.6$~per cent for the 0.66~$\rm M_\odot$ scenario and $2.1 \pm 0.7$~per cent for the fiducial simulation. This suppression is not simply the result of redistribution of already-formed HJ systems between subsamples. Rather, the absolute HJ formation rate across the entire simulation (combining both bound and escaped systems) decreases from 10.1~per cent in the fiducial model to 9.4~per cent in the 0.66~$\rm M_\odot$ scenario and further down to 8.1~per cent in the 1.4~$\rm M_\odot$ scenario. Amplified stochastic heating and velocity diffusion eject host stars from the cluster, while the stronger perturbations from close encounters ionize outer stellar companions before high-eccentricity migration can be completed. Consistently, roughly 71~per cent of systems in the 1.4~$\rm M_\odot$ model escape the cluster, compared to $\sim$18~per cent in the 0.66~$\rm M_\odot$ model and $\sim$18~per cent in the fiducial model. Among the escaping population, the HJ occurrence rate is correspondingly slightly elevated, reaching $1.3 \pm 0.4$~per cent for $m_{\rm pert} = 1.4~{\rm M_\odot}$ compared to $0.8 \pm 0.3$~per cent in the field sample.

We caution that the fixed perturber mass was applied uniformly across all cluster radii and simulation times, meaning that some systems in the 1.4~$\rm M_\odot$ model are artificially ejected by strong encounters far from the core where such perturbers would not realistically dominate. To resolve this, we ran a third experiment using a spatially dependent perturber mass to reflect a strongly segregated cluster. The radial dependence of the perturber mass was modelled as a step function, with a fixed $m_{\rm pert}$ set to 1.4~$\rm M_\odot$ for $r < r_{\rm h}$ and 0.66~$\rm M_\odot$ for $r \geq r_{\rm h}$, with the local number density evaluated correspondingly to conserve $\rho_\star$ and the close binary fraction set to $f_{\rm CB} = 0$. Similarly to the 1.4~$\rm M_\odot$ experiment, we find that the escape fraction in this scenario remains high at roughly 65~per cent. However, while the absolute global HJ formation rate further decreases to 7.4~per cent, the HJ yield in the cluster core reaches 11.8~per cent (compared to 11.1~per cent in the fiducial model). Correspondingly, the predicted HJ occurrence rate in the core remains elevated at $2.0 \pm 0.7$~per cent, closely matching our fiducial model.

The relative robustness of $\mathcal{F}_{\rm HJ}$ to the properties of the perturber population reflects our methodological choice to conserve the local mass density $\rho_\star$ across our experimental cluster setups. This behaviour can be understood based on simple yet illustrative analytical considerations. While the rate of strongly focused and therefore highly disruptive encounters scales with the mass of the perturber (Equation \ref{eq:gamma_focused}), that of more distant encounters which gently diffuse the system into the ZLK window operates in the geometric limit (Equation \ref{eq:gamma_geo}). In order to sufficiently perturb the eccentricity and inclination of the outer component of a 1P1S-type system to trigger ZLK cycles and subsequent tidal migration, a flyby must impart some threshold differential velocity kick $\Delta v_{\rm min}$ between the binary components. In the impulse approximation, the maximum impact parameter capable of delivering such a kick scales approximately as $b_{\rm max}^2 \propto m_{\rm pert} / (\Delta v_{\rm min} v_\infty)$. The geometric encounter rate of flybys capable of triggering HEM therefore becomes $\Gamma_{\rm HEM} \simeq n_\star \pi b_{\rm max}^2 v_\infty \propto n_\star m_{\rm pert} / \Delta v_{\rm min}$. Here $\rho_\star = n_\star m_{\rm pert}$, so the rate of HEM-triggering flybys scales as $\Gamma_{\rm HEM} \propto \rho_\star$. Hence, when varying the perturber mass $m_{\rm pert}$ while adjusting $n_\star$ to keep $\rho_\star$ constant, the resulting trade-off between the frequency and strength of close encounters conserves the total secular heating experienced by planetary systems over the cluster lifetime, leaving the rate of HJ formation largely unchanged.

While this artificial setup makes quantifying the exact effect of mass segregation on the HJ yield less straightforward, the results of these experiments suggest that a heavily segregated core does not dramatically increase the formation rate but rather primarily regulates the escape fraction. Therefore, while our fiducial simulation may underestimate the number of systems that escape the cluster, this only has limited implications for the predicted HJ occurrence rate in the cluster core. It should also be noted that the disruptive effect of high-mass perturbers may be further accentuated by a non-zero binary fraction which we explicitly omitted in these experiments in order to isolate the effect of the perturber mass. Ultimately, the precise impact of mass segregation depends on the timescale over which massive stars migrate to the core (related to the cluster relaxation time) relative to the timescale for HJ formation, a more detailed treatment of which we defer to future work.

\section{Self-consistency of Cluster Model}
\label{sec:selfconsistency}

\begin{figure}
\includegraphics[width=\columnwidth]{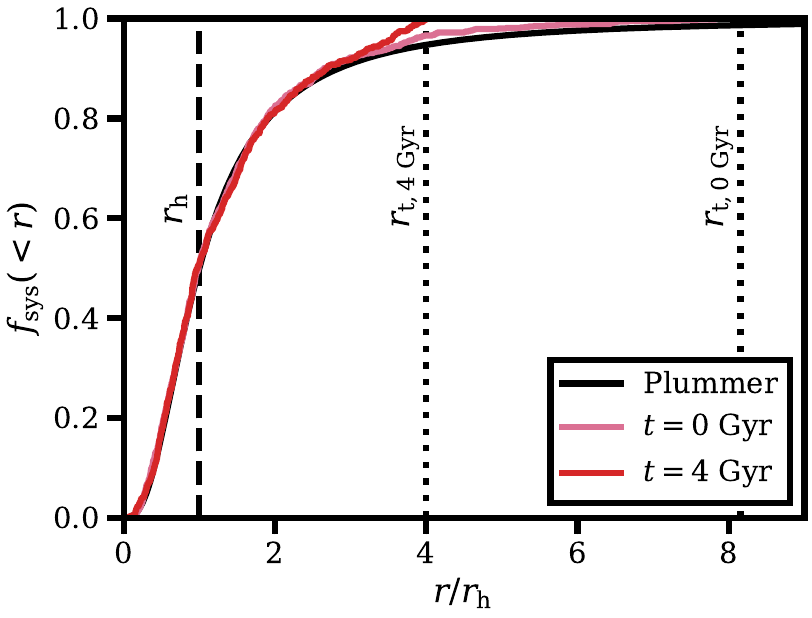}
\caption{The radial distribution of host stars matching the mean perturber mass (${\sim} 0.66~{\rm M_\odot}$) in the cluster at the beginning (mauve) and end (red) of the simulation. The good match with the underlying Plummer profile (black) supports the self-consistency of the model.
\label{fig:host_radii}}
\end{figure}

\begin{figure}
\includegraphics[width=\columnwidth]{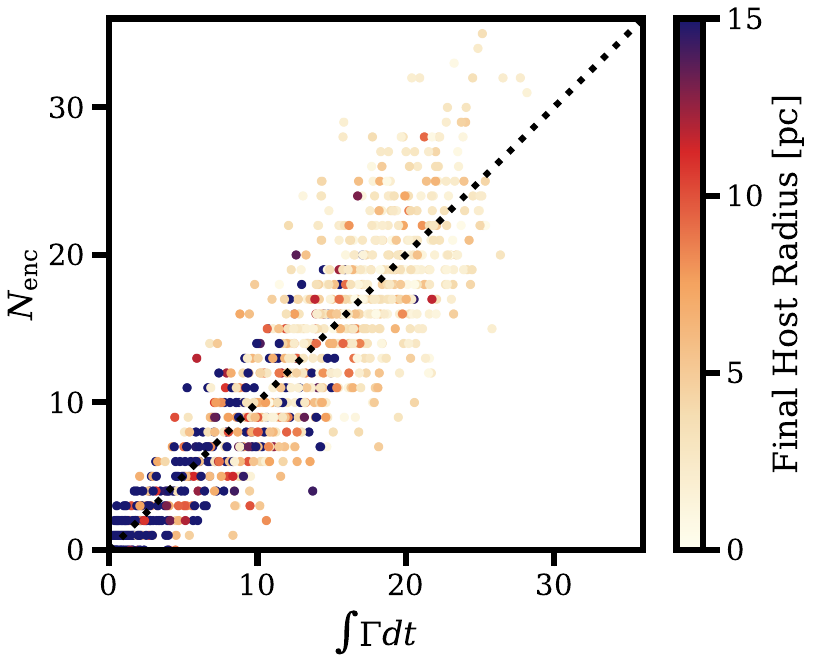}
\caption{The realized number of close encounters ($N_{\rm enc}$) within $r_{\rm enc}$ (Equation \ref{eq:renc}) against the integrated expectation value ($\int\Gamma dt$) for 1,000 systems with no planetary or stellar companions. The 1:1 correlation (dotted black line) confirms the self-consistency of the Poisson sampling algorithm. The marker colour denotes the final cluster-centric radius in parsecs, reflecting the scaling of the cumulative encounter rate with the local stellar density.
\label{fig:nencounters}}
\end{figure}

To test the self-consistency of our cluster evolution model, we initialize 1,000 host stars without planetary or stellar companions and evolve their trajectories in the cluster for $\sim$4~Gyr. The radius of the encounter sphere $r_{\rm enc}$ is then solely determined by the instantaneous strong encounter radius $r_{\rm str}$ (Equation \ref{eq:rstr}). For this experiment, we set the host star mass to $m_{\rm s} = \langle m_{\rm sys} \rangle \approx 0.66~{\rm M_\odot}$, such that the final radial distribution is expected to trace the underlying Plummer profile. As shown in Figure \ref{fig:host_radii}, the radial distribution of the host stars is in good agreement with the cluster as a whole at the end of the simulation. The deviation close to the final tidal radius is a result of the underlying potential in \texttt{gala} not being truncated at $r_{\rm t}$.

As a further verification of our stochastic encounter sampling framework, we compare the total realized number of close encounters ($N_{\rm enc}$) for each system against its cumulative expected encounter rate ($\int\Gamma dt$) integrated along the stellar trajectory over the $\sim$4~Gyr simulation. As shown in Figure \ref{fig:nencounters}, the number of realized encounters tracks the 1:1 expectation line with a scatter consistent with the variance of a Poisson process. Furthermore, the total number of encounters is strongly correlated with the final cluster-centric radius of the host star. Systems retained within the dense cluster core (red) predictably accumulate the highest number of encounters, whereas systems residing in the halo or escaping the cluster (blue) experience very few, confirming that our local sampling algorithm correctly captures the global density profile of the evolving cluster.

\section{Formulae for Eccentricity and Inclination Excitation}
\label{sec:formulae}

\subsection{Eccentricity Excitation}

Following \citet{heggie_effect_1996}, the eccentricity excitation experienced by a bound orbit $\{a,\:e\}$ due to a hyperbolic encounter in the secular limit is given by
\begin{equation}
\begin{split}
    \delta e = &-\frac{15}{4} \frac{e\sqrt{1-e^2}}{(1+e_{\rm pert})^{3/2}} 
    \frac{m_{\rm pert}}{\sqrt{m_{\rm ps} m_{\rm tot}}}
    \left(\frac{a}{r_{\rm close}}\right)^{3/2}
    \\ &\times \left[ \Theta_1 \chi + \left(\Theta_2 + \Theta_3\right)\psi \right]
\end{split}
\end{equation}
where
\begin{align}
    \chi &= \arccos\left(-\frac{1}{e_{\rm pert}}\right) + \sqrt{e_{\rm pert}^2 - 1} \\
    \psi &= \frac{(e_{\rm pert}^2 - 1)^{3/2}}{3e_{\rm pert}^2}
\end{align}
and the angular dependence is encoded in
\begin{align}
    \Theta_1 &= \sin^2{i}\sin{2\Omega} \\
    \Theta_2 &= (1 + \cos^2{i})\cos{2\omega}\sin{2\Omega} \\
    \Theta_3 &= 2\cos{i}\sin{2\omega}\cos{2\Omega}
\end{align}
where $\Omega$, $\omega$, and $i$ are the longitude of ascending node, argument of periapsis, and inclination of the perturber's trajectory relative to the bound orbit, $m_{\rm ps}$ is the total mass of the planetary system, $m_{\rm tot} = m_{\rm ps} + m_{\rm pert}$, and $r_{\rm close}$ is the closest approach distance of the perturber.

\subsection{Inclination Excitation}

Following \citet{rodet_odea_2019}, the inclination excitation experienced by a bound orbit $\{a,\:e\}$ due to a hyperbolic encounter in the secular limit is given by
\begin{equation}
\begin{split}
    \delta i = &\frac{3}{2} \frac{1 + 4e^2}{\sqrt{1 - e^2}\sqrt{1+e_{\rm pert}}} \frac{m_{\rm pert}}{\sqrt{m_{\rm ps} m_{\rm tot}}} 
    \left(\frac{a}{r_{\rm close}}\right)^{3/2}
    \\ &\times \left[ \Theta_4 \chi 
    - \Theta_5 \psi \right]
\end{split}
\end{equation}
where
\begin{align}
    \Theta_4 &= \cos{i}\sin{\Omega} \\
    \Theta_5 &= \sin{2\omega}\cos{\Omega} + \cos{i}\cos{2\omega}\sin{\Omega}
\end{align}
and the auxiliary functions $\chi$ and $\psi$ are as defined above.

\FloatBarrier

\bsp
\label{lastpage}
\end{document}